\documentclass[prc, reprint, amsmath, amssymb, superscriptaddress, nofootinbib, showpacs]{revtex4-1}

\usepackage{graphicx}%
\usepackage{multirow}
\usepackage{float}
\usepackage{epstopdf}
\usepackage[utf8]{inputenc}
\usepackage[english]{babel}
\usepackage{footnote}
\usepackage{xcolor}

\begin{document}

\title{Precise Q value measurements of $^{112,113}$Ag and $^{115}$Cd with the Canadian Penning trap for evaluation of potential ultra-low Q value $\beta$-decays}%

\author{N. D. Gamage}
\affiliation{Department of Physics, Central Michigan University, Mount Pleasant, MI 48859, USA}
\affiliation{National Superconducting Cyclotron Laboratory, East Lansing, Michigan, 48824, USA}
\email{Corresponding author: redsh1m@cmich.edu}

\author{R. Sandler}%
\affiliation{Department of Physics, Central Michigan University, Mount Pleasant, MI 48859, USA}

\author{F. Buchinger}
\affiliation{Department of Physics, McGill University, Montr\'{e}al, Qu\'{e}bec H3A 2T8, Canada}

\author{J. A. Clark}
\affiliation{Physics Division, Argonne National Laboratory, Argonne, IL 60439, USA}
\affiliation{Department of Physics and Astronomy, University of Manitoba, Winnipeg, Manitoba R3T 2N2, Canada}

\author{D. Ray}
\affiliation{Department of Physics and Astronomy, University of Manitoba, Winnipeg, Manitoba R3T 2N2, Canada}
\affiliation{Physics Division, Argonne National Laboratory, Argonne, IL 60439, USA}

\author{R. Orford}
\altaffiliation{Current address: Nuclear Physics Division, Lawrence Berkeley National Laboratory, Berkeley, CA 94720, USA}
\affiliation{Department of Physics, McGill University, Montr\'{e}al, Qu\'{e}bec H3A 2T8, Canada}
\affiliation{Physics Division, Argonne National Laboratory, Argonne, IL 60439, USA}

\author{W. S. Porter}
\affiliation{Department of Physics, University of Notre Dame, Notre Dame, IN 46556, USA}

\author{M. Redshaw}
\affiliation{Department of Physics, Central Michigan University, Mount Pleasant, MI 48859, USA}
\affiliation{National Superconducting Cyclotron Laboratory, East Lansing, Michigan, 48824, USA}

\author{G. Savard}
\affiliation{Physics Division, Argonne National Laboratory, Argonne, IL 60439, USA}
\affiliation{Department of Physics, University of Chicago, Chicago, IL 60637, USA}

\author{K.S. Sharma}
\affiliation{Department of Physics and Astronomy, University of Manitoba, Winnipeg, Manitoba R3T 2N2, Canada}

\author{A. A. Valverde}
\affiliation{Department of Physics, University of Notre Dame, Notre Dame, IN 46556, USA}
\affiliation{Physics Division, Argonne National Laboratory, Argonne, IL 60439, USA}

\date{\today}%

\begin{abstract}
\begin{description}
\item[Background]
An ultra-low Q value $\beta$-decay can occur from a parent nuclide to an excited nuclear state in the daughter such that $Q_{UL}$ $\lesssim$ 1 keV. These decay processes are of interest for nuclear $\beta$-decay theory and as potential candidates in neutrino mass determination experiments. To date, only one ultra-low Q value $\beta$-decay has been observed---that of $^{115}$In with $Q_\beta$ = 147(10) eV. A number of other potential candidates exist, but improved mass measurements are necessary to determine if these decay channels are energetically allowed and, in fact, ultra-low.
\item[Purpose]
To perform precise $\beta$-decay Q value measurements of $^{112,113}$Ag and $^{115}$Cd and to use them in combination with nuclear energy level data for the daughter isotopes $^{112,113}$Cd and $^{115}$In to determine if the potential ultra-low Q value $\beta$-decay branches of $^{112,113}$Ag and $^{115}$Cd are energetically allowed and $\lesssim$ 1 keV.
\item[Method]
The Canadian Penning Trap at Argonne National Laboratory was used to measure the cyclotron frequency ratios of singly-charged $^{112,113}$Ag and $^{115}$Cd ions with respect to their daughters $^{112,113}$Cd and $^{115}$In. From these measurements, the ground-state to ground-state $\beta$-decay Q values were obtained.
\item[Results]
The $^{112}$Ag $\rightarrow$ $^{112}$Cd, $^{113}$Ag $\rightarrow$ $^{113}$Cd, and $^{115}$Cd $\rightarrow$ $^{115}$In $\beta$-decay Q values were measured to be $Q_{\beta}$($^{112}$Ag) = 3990.16(22) keV, $Q_{\beta}$($^{113}$Ag) = 2085.7(4.6) keV, and $Q_{\beta}$($^{115}$Cd) = 1451.36(34) keV. These results were compared to energies of excited states in $^{112}$Cd at 3997.75(14) keV, $^{113}$Cd at 2015.6(2.5) and 2080(10) keV, and $^{115}$In at 1448.787(9) keV, resulting in precise $Q_{\textrm{UL}}$ values for the potential decay channels of --7.59(26) keV, 6(11) keV, and 2.57(34) keV, respectively.
\item[Conclusion]
The potential ultra-low Q value decays of $^{112}$Ag and $^{115}$Cd have been ruled out. $^{113}$Ag is still a possible candidate until a more precise measurement of the 2080(10) keV, 1/2$^{+}$ state of $^{113}$Cd is available. In the course of this work we have found the ground state mass of $^{113}$Ag reported in the 2020 Atomic Mass Evaluation [Wang, \textit{et al.}, Chin. Phys. C \textbf{45}, 030003 (2021)] to be lower than our measurement by 69(17) keV (a 4$\sigma$ discrepancy).
\end{description}
\end{abstract}

\maketitle

\section{Introduction}\label{Section_Intro}

Nuclear $\beta$-decays offer insight into the underlying weak interaction processes that govern them, and the in-medium effects that modify them due to their occurrence inside the atomic nucleus~\cite{Severijns2006,Haaranen2016_SSM}. The majority of unstable nuclides known to exist decay via allowed $\beta$-decay and have relatively large Q values. This results in them having typically short lifetimes, and making them fairly straight-forward to observe. However, there are a number of isotopes with low $\beta$-decay Q values and/or high forbiddenness, which results in them having much longer half-lives. These isotopes are important tools for applications such as direct neutrino mass determination experiments e.g.~\cite{Aker2021,Gas2017,Alp2015,Arnaboldi2003}, and radioactive dating e.g. \cite{Chechev2011}. It is also important to categorize and understand these rare decays since they can contribute to backgrounds in other rare event experiments, such as neutrinoless double $\beta$-decay (0$\nu\beta\beta$) and dark matter searches~\cite{Suhonen2018}. They also provide a testing ground for nuclear theory under atypical conditions~\cite{Mustonen2010}.

Ground-state to ground-state (gs--gs) $\beta$-decays have Q values covering a wide energy range from $\approx$2.5 keV up to $\sim$10\ MeV. However, under special circumstances, such as a decay from the ground state of the parent nuclide to a nearby excited state in the daughter, the Q value can be much lower. Such a decay, with $Q_{\textrm{UL}}$ $\lesssim$ 1 keV, is known as an ultra-low (UL) Q value $\beta$-decay~\cite{mus2010}.

To date, the only known UL Q value $\beta$-decay is that of the $^{115}$In(9/2$^{+}$) ground state to the $^{115}$Sn(3/2$^{+}$) first excited state. This decay was discovered by Cattadori, \textit{et al.} at the Laboratori Nazionali del Gran Sasso in 2005~\cite{cat2005}, and was later confirmed at the HADES underground laboratory in 2009~\cite{wie2009,And2011}. In these experiments, the $^{115}$In(9/2$^{+}$) $\rightarrow$ $^{115}$Sn(3/2$^{+}$) $\beta$-decay was inferred via the detection of the 497.5 keV $\gamma$-ray emitted from the $t_{1/2}$ = 11 ps, 3/2$^{+}$ state. Precise Penning trap measurements of the $^{115}$In -- $^{115}$Sn mass difference by groups at Florida State University~\cite{Mount2009} and the University of Jyv\"askyl\"a~\cite{Wieslander2009}, combined with the precisely known energy of the $^{115}$Sn(3/2$^{+}$) state, showed that this decay is energetically allowed with a $Q_{\textrm{UL}}$ value of 147(10)
eV \footnote{Here we use Q$_{\beta}$ = 497.489(10) keV from Ref.~\cite{Wan2017} and E[$^{115}$Sn(3/2$^{+}$)] = 497.342(3) keV from the recent measurement of Ref.~\cite{Zhe2019}.}. Theoretical descriptions of this UL Q value decay showed significant discrepancies between the calculated and measured partial half-life~\cite{mus2010,Suh2010}. The identification of additional UL Q value decays, and measurements of their partial half-lives are required to aid further theoretical developments. Furthermore, UL Q value $\beta$-decays have the potential to be new candidates for direct neutrino mass determination experiments, since the fraction of decays in the energy interval $\Delta E$ close to the endpoint, which is relevant for the determination of the neutrino mass, goes as $(\Delta E/Q)^{3}$~\cite{Ferri2015}.

Potential UL Q value decay branches were identified in $^{115}$Cd~\cite{Haa2013} and $^{135}$Cs~\cite{Mus2011} with $Q_{\textrm{UL}}$ values of -2.8(4.0) keV and 0.5(1.1) keV respectively, limited by the uncertainties in the masses of the parent and/or daughter isotopes. Since Ref.~\cite{Haa2013} was published, new atomic mass data in the 2016 Atomic Mass Evaluation (AME2016)~\cite{Wan2017} gave an UL Q value for $^{115}$Cd of 3.1(0.7) keV\footnote{This Q value remains the same in the most recent 2020 update to the Atomic Mass Evaluation (AME2020)~\cite{Wang2021}.}. Although this would indicate that this transition is not UL, we note that in AME2016 and AME2020 the $^{115}$Cd mass is determined entirely through a ($d,p$) reaction measurement linking it to $^{114}$Cd. Since atomic masses obtained via nuclear reaction data are not always reliable, and because the mass of $^{114}$Cd was determined from an older mass spectrometry technique, a direct Penning trap measurement of the $^{115}$Cd Q value is called for.
Recently, the ground-state Q value of $^{135}$Cs was measured with the JYFLTRAP Penning trap and the Q value of the potential UL decay branch was determined to be $Q_{\textrm{UL}}$ = 0.44(31) keV~\cite{deRoubin2020}, showing that it is indeed energetically allowed with $Q_{\textrm{UL}} <$ 1 keV. 

Additional potential UL Q value decay candidates have been identified in the literature~\cite{mus2010_2,Suh2014,kop2010,gam2019}. Again, more precise mass data for the parent and/or daughter isotopes are required to determine whether these decays are energetically allowed and if their Q values are ultra-low. In collaboration with the LEBIT group at the National Superconducting Cyclotron Laboratory, we recently investigated and ruled out two candidates identified in Ref.~\cite{gam2019}: $^{89}$Sr and $^{139}$Ba~\cite{Sandler2019_89Y}. The JYFLTRAP group recently ruled out potential UL Q value decays in $^{72}$As~\cite{Ge2021}, and showed that $^{159}$Dy does have an electron capture decay branch to an excited state in the daughter with a $Q_{\textrm{UL}}$ value of around 1 keV~\cite{Ge2021_159Dy}. They are also investigating other candidates~\cite{Kan2020}.

\begin{figure}[ht]
    \includegraphics[width=0.85\columnwidth]{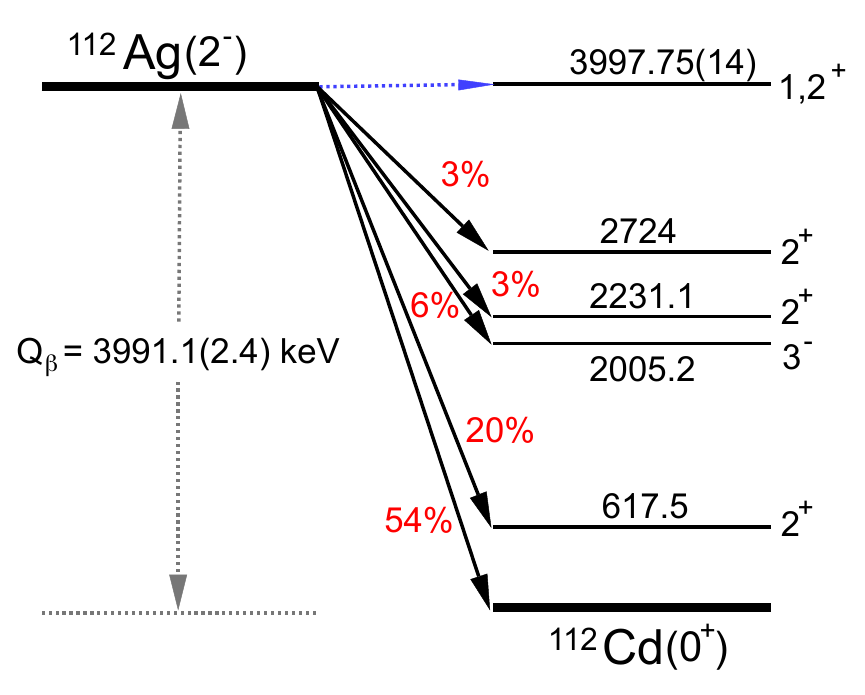}
    \includegraphics[width=0.85\columnwidth]{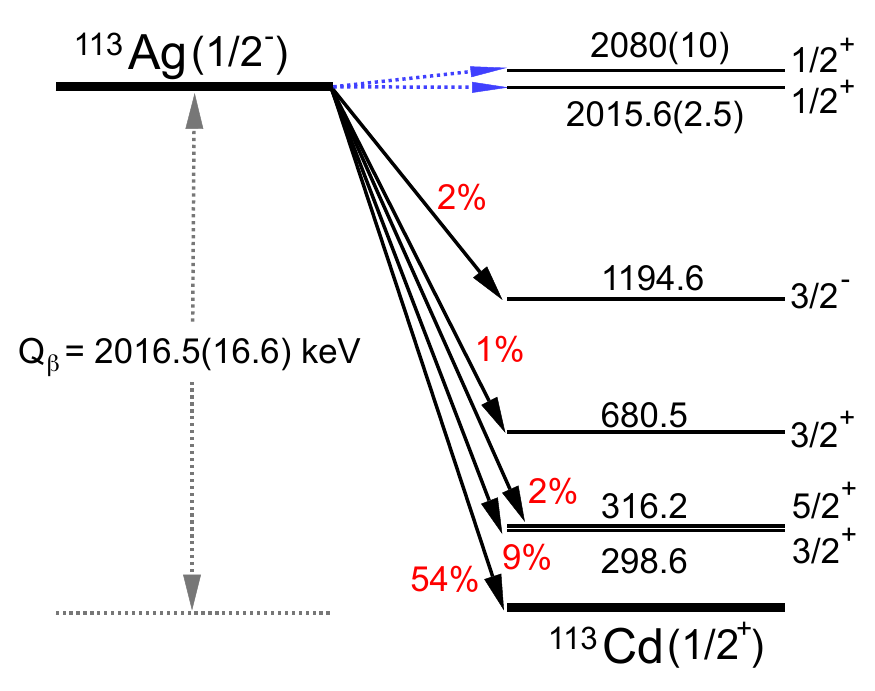}
    \includegraphics[width=0.85\columnwidth]{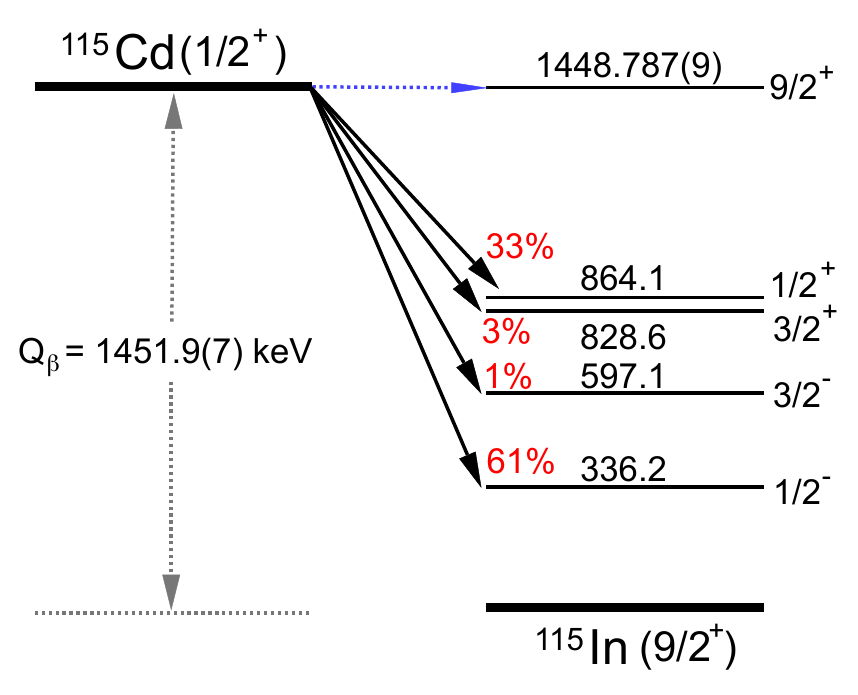}
    \caption{(color online) Decay schemes for $^{112,113}$Ag and $^{115}$Cd showing the main $\beta$-decay branches (solid black arrows) and the potential ultra-low Q value decay branches (dashed blue arrows) investigated in this work. The ground-state to ground-state Q values are obtained using data from the AME2020~\cite{Wang2021}. All values are given in units of keV.}
    \label{fig:Level_Scheme}
\end{figure}

In this paper we present the first direct measurement of the $^{115}$Cd gs--gs Q value ($Q_{gs}$) and determination of the Q value of the potential UL decay branch identified in Ref.~\cite{Haa2013}. We also investigate two potential UL Q value candidates identified in Ref.~\cite{gam2019}: $^{112,113}$Ag. Decay schemes for all three candidates are shown in Fig. \ref{fig:Level_Scheme}. 
For $^{112}$Ag and $^{115}$Cd, potential UL Q value decays are to the 1 or 2$^+$ level at 3997.8 keV in $^{112}$Cd, and to the 9/2$^+$ level at 1448.8 keV in $^{115}$In, respectively. In the case of $^{113}$Ag, two potential UL Q value branches to 1/2$^+$ states in $^{113}$Cd at 2015.6 keV and 2080 keV are shown. A decay to the lower energy state was identified as a potential UL Q value transition based on the gs--gs Q value obtained using mass data from AME2016~\cite{Wan2017}, but was ruled out by our measurement presented here. Based on our new result, a decay to the higher energy state was later identified as a potential UL Q value transition.

\section{Experimental Description}\label{Meth_Section}
The $^{112,113}$Ag, and $^{115}$Cd gs--gs Q values were obtained via measurements of the cyclotron frequency ratio of singly-charged ions of the parent and daughter isotopes using the Canadian Penning Trap (CPT) at Argonne National Laboratory (ANL)~\cite{savard2001,orford2020}.

CPT is currently coupled to the Californium Rare Isotope Breeder Upgrade (CARIBU) facility that produces neutron-rich rare isotopes via spontaneous fission of $^{252}$Cf~\cite{savard2008}. Fission products are thermalized in a gas catcher~\cite{SAVARD2003} and extracted as singly- or doubly-charged ions. The ions are then accelerated and go through a high-resolution mass separator to select ions of a particular $A/q$. The selected ions then enter into a radio-frequency quadrupole (RFQ) cooler and buncher to prepare low emittance bunches for injection into the CARIBU multi-reflection time-of-fight mass separator (MR-TOF-MS)~\cite{HIRSH2016} where the next stage of mass selection occurs as ions reflect between electrostatic mirrors and separate in time-of-flight with respect to their mass, providing typical mass resolving powers of up to 100,000. A Bradbury-Nielsen Gate (BNG)~\cite{BNG1936} is then used to select a particular isotope. However, depending on the mass difference between isobars, more than one species can pass through the BNG. Finally,  ion bunches are accumulated and further cooled in a linear Paul trap before being injected into the Penning trap.

The CPT has a hyperbolic geometry with compensation ring and tube electrodes and is immersed in a uniform 5.9 T magnetic field produced by a superconducting solenoidal magnet. Ions confined in the Penning trap undergo three normal modes of motion: axial, reduced-cyclotron, and magnetron, with characteristic frequencies $f_z$, $f_+$ and $f_-$, respectively~\cite{Brown1986}. By combining measurements of the observable normal mode frequencies, one can obtain the free-space cyclotron frequency for an ion with charge-to-mass ratio $q/m$ in a magnetic field of strength $B$:
\begin{equation}\label{CycFreqEqn}
    f_c = \frac{qB}{2\pi m}.
\end{equation}
At the CPT, $f_c$ is measured using the so-called Phase-Imaging Ion Cyclotron Resonance (PI-ICR) technique that was originally developed and implemented by the SHIP-TRAP group~\cite{eliseev2013,eliseev2014}. This technique enables a measurement of the total phase accumulated by an ion in its reduced-cyclotron or magnetron motion during a precisely defined time interval, that in turn can be used to determine the ion's frequency in the trap. The phase determination is performed by ejecting ions from the trap onto a position sensitive micro channel plate (MCP) detector, preserving the ion's phase information.

In this work, the direct method for determining $f_c$ described in Ref.~\cite{eliseev2014} was used~\cite{orford2020}. This method involves two separate phase measurements known as the reference spot measurement and the final spot measurement (an additional measurement at the start of the experiment is also required to determine the spot on the MCP that corresponds to the center of the trap). For both the reference and final spot measurements, the reduced-cyclotron motion of the ion is first excited to a well-defined radius via a pulsed rf dipole drive at a frequency close to $f_+$. For the reference spot, a quadrupole rf drive pulse at frequency $f_{\textrm{rf}} \approx f_{+} + f_{-} = f_c$ is immediately applied, which converts the ion's reduced-cyclotron motion into magnetron motion. The ion is then allowed to accumulate phase in its magnetron motion for a specific time period before being ejected from the trap. For the final spot, the quadrupole rf drive pulse is applied after a phase accumulation period of length $t_{\textrm{acc}}$, so that the ion accumulates mass-dependent reduced-cyclotron phase before its reduced-cyclotron motion is converted into magnetron motion. The ion then remains in its magnetron orbit for an additional period of time until it is ejected such that the total time spent in the trap during the reference and final spot measurements is the same. The cyclotron frequency, $f_{c}$, is determined from the total phase difference $\Delta\phi$ between the reference and measurement spots during the time interval $t_{\textrm{acc}}$ \cite{eliseev2014}
\begin{equation}
    f_c = \frac{\Delta\phi}{2\pi t_{\textrm{acc}}} = \frac{\phi_{\textrm{meas}} + 2\pi N}{2\pi t_{\textrm{acc}}},
    \label{Eqn_PI-ICR_Freq}   
\end{equation}
where $\phi_{\textrm{meas}}$ is the measured angle between the reference and final spot and $N$ is the number of complete revolutions for an ion with cyclotron frequency $f_{c}$ during time $t_{\textrm{acc}}$.

\begin{figure}[ht]
\includegraphics[width=\columnwidth]{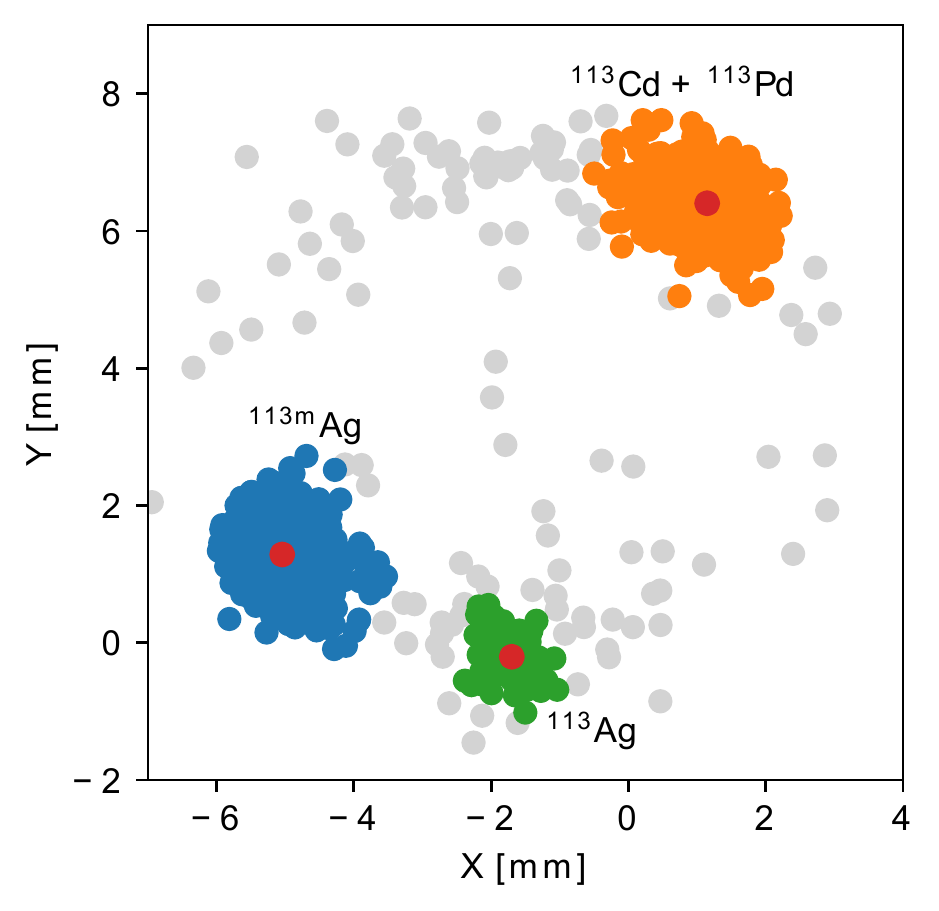}
\caption{(Color online) An example of a PI-ICR data set showing the separation of $^{113}$Ag (green spot) and the $^{113m}$Ag isomer (blue spot), as well as $^{113}$Cd and $^{113}$Pd (orange spot) isobaric contaminants, as identified with the cluster-finding algorithm. The red dot in each spot shows the center of each cluster. Grey dots represent counts on the MCP that are identified as noise. In this data set, a 550 ms $t_{\textrm{acc}}$ was used.}\label{Fig:113Ag_PI-ICR} 
\end{figure}

In order to obtain $\phi_{\textrm{meas}}$, the central coordinates and associated uncertainties of the reference spot and final spot need to be determined. This was done using an unsupervised learning cluster-finding model, the Gaussian mixture model, which has been developed based on an expectation–maximization algorithm, see e.g. Ref.~\cite{Weber2022}. For the final phase measurement, there can be several spots along with the spot corresponding to the ion of interest depending on contaminants present in the beam. In such cases, $t_{\textrm{acc}}$ was carefully chosen so that the spot of interest is well separated from the other spots. Such an example can be seen in Figure~\ref{Fig:113Ag_PI-ICR}, which displays the output of the cluster-finding model and shows the presence of $A$ = 113 isobars $^{113}$Ag, $^{113\textrm{m}}$Ag, $^{113}$Cd, and $^{113}$Pd. Typically $\sim$100 -- 500 ions were accumulated in a spot to enable a determination of $f_{c}$ to a precision of $\sim$3 -- 5 mHz. Such a measurement took $\sim$2 -- 30 mins, depending on the rate at which the isotope of interest was delivered to the Penning trap. If necessary, this rate was limited to allow only a few ions per shot into the trap to avoid potential systematic frequency shifts due to ion-ion interactions.

\begin{figure}[ht]
\includegraphics[width=\columnwidth]{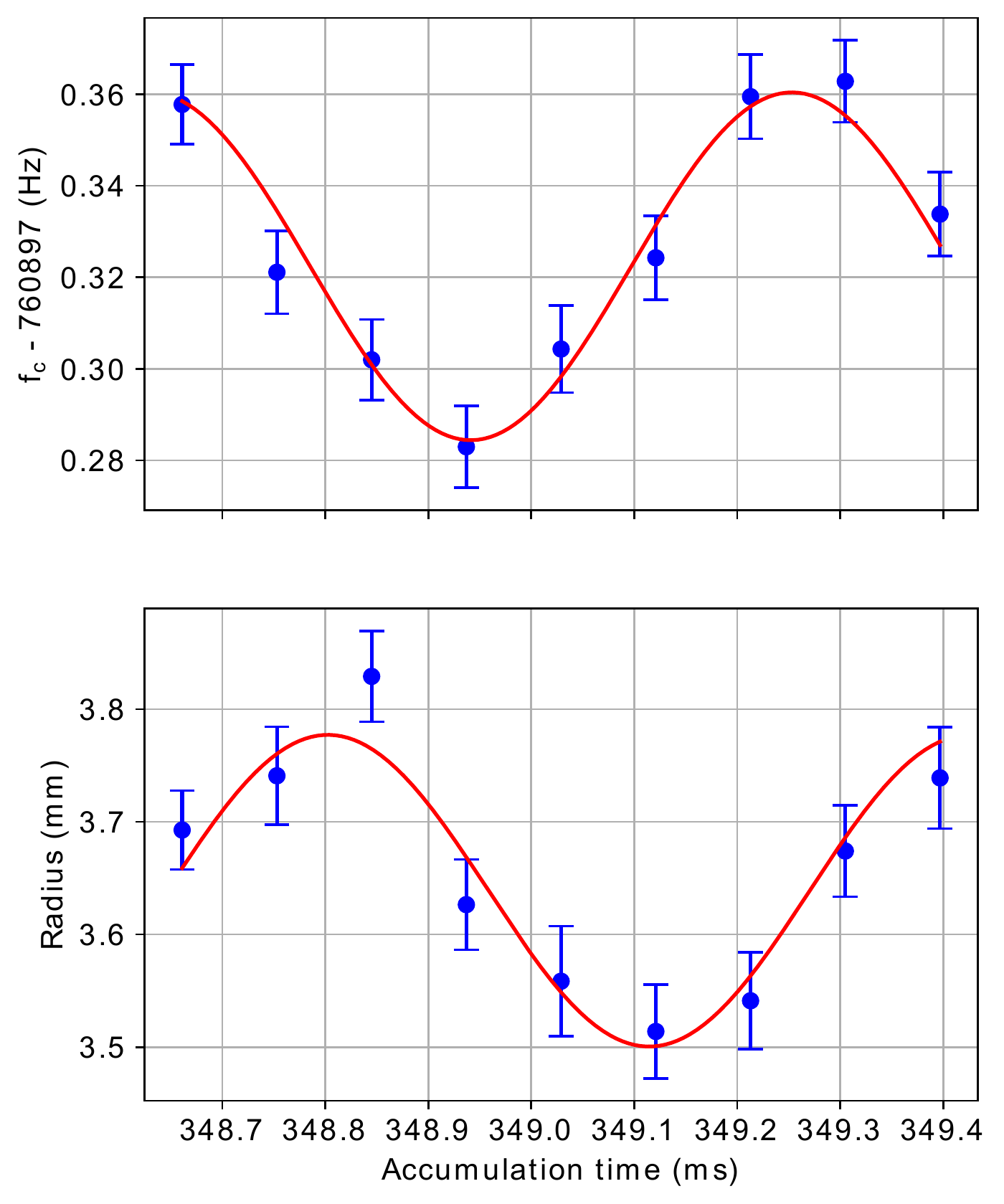}
\caption{(Color online) Sinusoidal variation of $f_c$ and radius of the projected orbit of the $^{115}$Cd ion spot observed on the MCP as a function of accumulation time, $t_{\textrm{acc}}$, from the 2017 data set.}
\label{113Cd_fc_and_R_vs_tacc} 
\end{figure}

\section{Data and Analysis}\label{Data_Section}
\subsection{Experimental Runs and Data}
The experimental data were acquired during three separate runs in 2016, 2017 and 2018. During the initial 2016 run, data were taken for the ratios of interest $^{112}$Ag$^{+}$/$^{112}$Cd$^{+}$, $^{113}$Ag$^{+}$/$^{113}$Cd$^{+}$, and $^{115}$Cd$^{+}$/$^{115}$In$^{+}$, and for test ratios $^{112}$Sn$^{+}$/$^{112}$Cd$^{+}$ and $^{115}$In$^{+}$/$^{115}$Sn$^{+}$ that involve isotopes whose masses have already been precisely measured with Penning traps~\cite{Mount2009,Wieslander2009,Rahaman2009}. In this data we observed shifts in the Q values calculated from the test ratio measurements of up to 10 keV compared to literature values. We also observed variations in the Q values for different $t_{\textrm{acc}}$ times that were found to result from variations in $f_{c}$ for different $t_{\textrm{acc}}$ times.

After the 2016 run, it was discovered that these shifts were due to a systematic sinusoidal variation of $f_{c}$ as a function of $t_{\textrm{acc}}$ with a frequency corresponding to the magnetron frequency of ions in the trap. A corresponding systematic variation was observed in the final radial position of the ions on the MCP with a 90$^{\circ}$ phase shift compared to the $f_{c}$ data, see Figure \ref{113Cd_fc_and_R_vs_tacc} and Ref.~\cite{orford2020}. These observations indicate that ions injected into the trap had some initial magnetron motion with a reproducible amplitude and phase when they were initially confined in the trap, before their reduced-cyclotron motion was first driven by the pulsed rf dipole drive. This motion is then transferred to the final magnetron motion of the ions before they are ejected from the trap, with a phase that depends on the phase accumulated during the $t_{\textrm{acc}}$ period. Hence, the position of the final spot on the MCP is modified slightly resulting in a final radial position and phase, $\phi_{\textrm{meas}}$, that depend on $t_{\textrm{acc}}$. Following this discovery, a second experimental run was performed in 2017 with the $A = 115$ ion pairs $^{115}$Cd$^{+}$/$^{115}$In$^{+}$ and $^{115}$In$^{+}$/$^{115}$Sn$^{+}$, and a third experimental run was undertaken in 2018 to take additional $A = 115$ data and data for the $A = 112$ ion pairs $^{112}$Ag$^{+}$/$^{112}$Cd$^{+}$ and $^{112}$Sn$^{+}$/$^{112}$Cd$^{+}$.

During the 2017 and 2018 runs, $t_{\textrm{acc}}$ was systematically varied to map out and account for the sinusoidal variation of $f_{c}$ versus $t_{\textrm{acc}}$. A sinusoidal fit to the data of the form
\begin{equation}
    f_c(t) = f_{c0} + A_0(t) \textrm{sin}(2\pi f _{-}t + \psi)
\label{Eqn_SineFit}    
\end{equation}
was then performed. In this fit, $A_{0}(t)$ is the amplitude of the sine fit function, $\psi$ is a phase offset, and $f_{c0}$ is the baseline cyclotron frequency when no systematic shift occurs. Hence, $f_{c0}$ and its associated uncertainty are extracted. In this fit, $f_{-}$ was constrained to the measured magnetron frequency. We note that, although the magnitude of the shift in $\phi_{\textrm{meas}}$ does not depend on $t_{\textrm{acc}}$, $A_{0}$ does because $f_c$ from Eqn. (\ref{Eqn_PI-ICR_Freq}) goes as 1/$t_{\textrm{acc}}$.  

The procedure to measure $f_{c0}$ in this way was then repeated with the other isotope for the parent-daughter pair so that the Q value could be obtained, as discussed in section \ref{Results_Section}. Depending on the isotope pair, between one and three $f_{c0}$ measurements were performed for each isotope, alternating between the two. 

After obtaining $f_{c0}$ for ions of the parent and daughter isotopes, the cyclotron frequency ratio, corresponding to the inverse mass ratio of the ions, was obtained:
\begin{equation}\label{fc_ratio_eqn}
R = \frac{f_{c0}^{p}}{f_{c0}^{d}} = \frac{m_{d}}{m_{p}}.
\end{equation}
In the case that more than one $f_{c0}$ measurement was performed for parent and/or daughter isotope, neighboring $f_{c0}$ measurements for one ion were linearly interpolated to the time of the $f_{c0}$ measurement of the other ion to account for temporal magnetic field drifts. A weighted average of all resulting cyclotron frequency ratio measurements for a given ion pair was then obtained. The average cyclotron frequency ratios, after applying the systematic corrections discussed below, are given in Table \ref{table:Ratios}. 

\begin{table}[]
\caption{\label{table:Ratios} Average cyclotron frequency ratios obtained from the measurements performed in this work. Measurements were performed in three separate experimental runs in 2016, '17 and '18. $\Delta R_{\mathrm{fin}}$ and $\Delta R_{\mathrm{ref}}$ are corrections (and associated uncertainties in parentheses) $\times$10$^{9}$ applied to the ratio to account for systematic shifts to the final and reference spots, respectively (see text for details). $\sigma_{\mathrm{st}}$ is the statistical uncertainty $\times$10$^{9}$, and $\bar{R}$ is the resulting corrected ratio, with combined statistical and systematic uncertainties in parentheses.}
\begin{ruledtabular}
\begin{tabular}{cccccl}
Ion Pair & Yr. & $\Delta R_{\mathrm{fin}}$ & $\Delta 
R_{\mathrm{ref}}$ & $\sigma_{\mathrm{st}}$ & \multicolumn{1}{c}{$\bar{R}$}\\
\hline
$^{112}\text{Ag}^+$/$^{112}\text{Cd}^+$ & `16 & 0(48) & 48(5) & 8 & 0.999 961 658(49) \\
$^{112}\text{Ag}^+$/$^{112}\text{Cd}^+$ & `18 & 0(0) & -1.6(2) & 2.1 & 0.999 961 721 5(21) \\
\hline

$^{112}\text{Sn}^+$/$^{112}\text{Cd}^+$ & `16 & 0(48) & 18(2) & 9 & 0.999 981 445(49) \\
$^{112}\text{Sn}^+$/$^{112}\text{Cd}^+$ & `18 & 0(0) & 2.1(2) & 2.4 & 0.999 981 582 2(24) \\
\hline

$^{113}$Ag$^{+}$/$^{113}$Cd$^{+}$ & `16 & 0(42) & 2.3(2) & 12 & 0.999 980 169(44) \\
$^{113}$Ag$^{+}$/$^{113\mathrm{m}}$Ag$^{+}$ & `16 & 0(41) & 7.5(8) & 15 & 0.999 999 616(44) \\
\hline

$^{115}\text{Cd}^+$/$^{115}\text{In}^+$ & `16 & -4(9) & 15(2) & 1.4 & 0.999 986 451 4(93) \\
$^{115}\text{Cd}^+$/$^{115}\text{In}^+$ & `17 &  0(0) & -19(2) & 6.0 & 0.999 986 435 5(63) \\
$^{115}\text{Cd}^+$/$^{115}\text{In}^+$ & `18 & 0(0) & -1.4(2) & 3.9 & 0.999 986 440 1(39) \\
 \multicolumn{5}{c}{Average} & 0.999 986 440 3(31) \\
\hline
$^{115}\text{In}^+$/$^{115}\text{Sn}^+$ & `16 & 26(13) & 5(1) & 2.2 & 0.999 995 361(13) \\
$^{115}\text{In}^+$/$^{115}\text{Sn}^+$ & `17 &  0(0) & 14(2) & 3.9 & 0.999 995 351 8(44) \\
$^{115}\text{In}^+$/$^{115}\text{Sn}^+$ & `18 & 0(0) & 4.9(5) & 7.0 & 0.999 995 332 7(70) \\
\end{tabular}
\end{ruledtabular}
\end{table}

\subsection{Systematic Corrections and Checks}

During the experimental runs in 2017 and 2018, the data were obtained by measuring $f_c$ as a function of $t_{\textrm{acc}}$ and extracting the baseline value, $f_{c0}$, from a fit using Eqn. (\ref{Eqn_SineFit}). This procedure enabled us to account for the phase dependent shift to $f_c$ as a function of $t_{\textrm{acc}}$ so that it did not affect the cyclotron frequency ratio, $R$. Hence there is no $\Delta R_{\mathrm{fin}}$ correction to the 2017 or 2018 data in Table \ref{table:Ratios}.

For the 2016 data, measurements were performed for each ratio at specific $t_{\textrm{acc}}$ times, resulting in frequencies and corresponding ratios that did suffer from systematic shifts. In the case of $^{115}$Cd$^{+}$/$^{115}$In$^{+}$, we were able to use the parameters from the fit of Eqn. (\ref{Eqn_SineFit}) to the 2017 $A = 115$ data to correct the 2016 data. The correction, $\Delta R_{\mathrm{fin}}$, and the corrected ratio are shown in Table \ref{table:Ratios}. After this correction, there is good agreement between the corrected 2016 $^{115}$Cd$^{+}$/$^{115}$In$^{+}$ data and the data taken in 2017 and 2018. As such, we averaged $^{115}$Cd$^{+}$/$^{115}$In$^{+}$ data from all three runs to obtain the average ratio listed in Table~\ref{table:Ratios} to be used to determine the $^{115}$Cd gs-gs $\beta$-decay Q value.

Correcting the 2016 $A = 115$ data was possible because the 2017 data was taken within two months of the 2016 data and no significant changes to the CPT apparatus were made. Before the 2018 run, some 18 months after the 2017 run, a magnet quench had occurred, requiring the magnet to be re-energized, and a new voltage source for the Penning trap electrodes had been installed. As such, the parameters from the fit of Eqn. (\ref{Eqn_SineFit}) to the 2018 data did not reproduce those from the 2017 data. Hence, we were not able to use the 2018 $A = 112$ data to correct the 2016 $^{112}$Ag$^{+}$/$^{112}$Cd$^{+}$ data. We also did not take additional data at $A = 113$ in 2018 because the 2016 data was sufficient to rule out the potential UL Q value decay branch. As such, we do not apply a correction, $\Delta R_{\mathrm{fin}}$, to the 2016 $^{112,113}$Ag data. Instead, we include an uncertainty due to this effect that we estimate from Eqn. (\ref{Eqn_PI-ICR_Freq}) and (\ref{Eqn_SineFit}) based on the $t_{\textrm{acc}}$ that was used and the variation in the observed orbital radius of the ion spot on the MCP, as seen, for example, in Figure (\ref{113Cd_fc_and_R_vs_tacc}).
We also applied the $\Delta R_{\mathrm{fin}}$ correction to the 2016 $^{115}$In$^{+}$/$^{115}$Sn$^{+}$ test ratio data, and included the systematic uncertainty due to this effect in the $^{112}$Sn$^{+}$/$^{112}$Cd$^{+}$ test ratio, as reported in Table II.

After the 2018 run, a smaller, additional systematic shift to $f_{c}$ data taken using the PI-ICR technique at the CPT was discovered~\cite{orford2020}. This shift affected the phase of the reference spot and is due to contaminant ions of the same nominal $A/q$ in the trap. 
During the pulsed rf dipole drive at $f_{+}$  that is used to initially drive the ions to their reduced cyclotron orbit, and also during the pulsed rf quadrupole drive at $f_{c}$ that is used to convert cyclotron motion into magnetron motion, ions accumulate a phase difference that depends on $m/q$. Because the $f_{+}$ and $f_{c}$ drive pulses are of short duration ($\approx$500\: $\mu$s), the resulting phase difference is typically small ($\sim$5$^\circ$), and a separation of different species in the reference spot is not observed. However, depending on the proportion of contaminant ions to the ion of interest entering the trap, the weighted average phase of the reference spot can be systematically offset from the phase of just the ions of interest~\cite{orford2020}.
This effect can be corrected for by determining the percentage of contaminant ions vs ions of interest and calculating the corresponding weighted phase shift. This correction was accounted for in the data, and has been included in Table \ref{table:Ratios} as the correction $\Delta R_{\textrm{ref}}$.

Our cyclotron frequency ratio measurements of $^{112}$Sn$^{+}$/$^{112}$Cd$^{+}$ and $^{115}$In$^{+}$/$^{115}$Sn$^{+}$ were performed to serve as an independent check of our measurements by comparing them to the inverse mass ratios calculated with data listed in the AME2020~\cite{Wang2021}. We also observed the $^{113\mathrm{m}}$Ag isomer, so were able to obtain the ratio $^{113\mathrm{m}}$Ag$^{+}$/$^{113}$Ag$^{+}$, which can be compared to the mass ratio of the $^{113}$Ag ground and isomeric state. A comparison of these data is shown in Table \ref{table:R_comparison}.

\begin{table}[]
\caption{\label{table:R_comparison} Comparison of test ratio measurements, $\bar{R}$, given in Table \ref{table:Ratios} to inverse mass ratios, $R_\mathrm{lit}$, obtained from literature values for mass \cite{Wang2021} and isomeric energies \cite{Kondev2021}. $\Delta R$ is the difference ($\bar{R}$ -- $R_\mathrm{lit}$) $\times$10$^{9}$ with total uncertainty in parentheses.}
\begin{ruledtabular}
\begin{tabular}{cclc}
Ion Pair & Year & \multicolumn{1}{c}{$R_{\mathrm{lit}}$} & \multicolumn{1}{c}{$\Delta R$ ($\times$10$^{-9}$)}\\
\hline
\multirow{2}{*}{$^{112}$Sn$^{+}$/$^{112}$Cd$^{+}$} & 2016 & \multirow{2}{*}{0.999 981 582 5(37)} & -137(49)\\
 & 2018 & & -0.3(4.4)\\
\hline
$^{113}\text{Ag}^+$/$^{113m}\text{Ag}^+$ & 2016 & 0.999 999 586 4(10) & 30(44)\\
\hline
 & 2016 & & 9(13)\\
 $^{115}\text{In}^+$/$^{115}\text{Sn}^+$ & 2017 &  0.999 995 351 9(2) & -0.2(4.4)\\
 & 2018 & & -19(7)\\
\end{tabular}
\end{ruledtabular}
\end{table}

\section{Results and Discussion}\label{Results_Section}
\subsection{$Q_{gs}$ values for $^{112,113}$Ag and $^{115}$Cd $\beta$-decay }\label{SubSection_Qvalues}

\begin{figure}[t]
\includegraphics[width=\columnwidth]{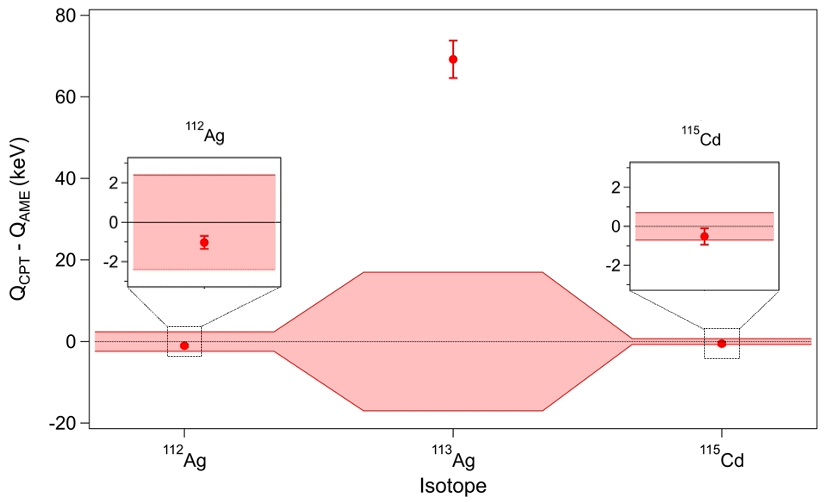}
\caption{(Color online) Ground-state to ground-state Q values measured in this work. The red bands show the AME2020 uncertainty and the red dots are our measured values.  \label{fig:QValues}} 
\end{figure}

The goal of this work was to obtain gs--gs $\beta$-decay Q values for $^{112,113}$Ag and $^{115}$Cd. These Q values, defined as the energy equivalent of the mass difference between parent and daughter atoms, can be determined from the measured cyclotron frequency ratios via
\begin{equation}\label{Eqn_Qgs}
    Q_{gs} = (M_p - M_d)c^2 = (M_d - m_e)(\bar{R}^{-1} - 1)c^2,
\end{equation}
where $M_p$ and $M_d$ are the mass of the parent and daughter atoms, respectively, and $m_e$ is the mass of the electron. The conversion factor from atomic mass units to keV, 1 u $= 931\,494.102\,42(28)$ keV/$c^2$ from Ref. \cite{CODATA2018} was used.
The average cyclotron frequency ratios listed in Table \ref{table:Ratios} were used to obtain these Q values, and the results are listed in Table \ref{table:Q}. Daughter atomic masses were taken from the AME2020 \cite{Wang2021}, and $m_e$ from the most recent CODATA recommended values of the fundamental physical constants \cite{CODATA2018}. Q values obtained in this work are compared to those from the AME2020 in Table \ref{table:Q} and in Fig. \ref{fig:QValues}.

Our Q values for $^{112}$Ag and $^{115}$Cd are slightly lower, but in agreement with the values obtained from the AME2020. Our result for $^{112}$Ag is a factor of 10 more precise, while that for $^{115}$Cd is about a factor of two more precise. Our result for $^{113}$Ag shows a significant, 4$\sigma$ discrepancy compared to the AME2020, and is a factor of three more precise.
\begin{table}[b]
 \caption{\label{table:Q} Q values obtained in this work, $Q_\mathrm{CPT}$, from the cyclotron frequency ratio measurements listed in Table~\ref{table:Ratios} and using Eqn.~(\ref{Eqn_Qgs}). Results obtained from the AME2020~\cite{Wang2021}, $Q_\mathrm{AME}$, and the differences $\Delta Q$ = $Q_\mathrm{CPT}$ -- $Q_\mathrm{AME}$ are also listed. All values are in keV.}.
\begin{ruledtabular}
\begin{tabular}{cccc}
Decay & $Q_\mathrm{CPT}$ & $Q_\mathrm{AME}$ & $\Delta Q$ \\
\hline
$^{112}$Ag $\rightarrow$ $^{112}$Cd & 3990.16(22) & 3991.1(2.4) & -1.0(2.4) \\
$^{113}$Ag $\rightarrow$ $^{113}$Cd & 2085.7(4.6) & 2016.5(16.6) & 69.2(17.3) \\
$^{115}$Cd $\rightarrow$ $^{115}$In & 1451.36(34) & 1451.88(65) & -0.52(73)\\
\end{tabular}
\end{ruledtabular}
\end{table}

\subsection{Evaluation of $Q_{\textrm{UL}}$ values for $^{112,113}$Ag and $^{115}$Cd}\label{SubSection_ULQvalues}

Using the new Q values listed in Table \ref{table:Q}, the potential UL Q values for decay branches to excited states in the daughter nuclei can be evaluated via
\begin{equation}\label{Eqn_UL_Qvalue}
  Q_{\textrm{UL}} = Q_{gs} - E^{*}, 
\end{equation}
where $Q_{gs}$ is the gs--gs Q value from Table \ref{table:Q}, and $E^{*}$ is the energy of the final state in the daughter, from Ref. \cite{nndc}. The $Q_{\textrm{UL}}$ values that we obtained are listed in Table \ref{table:QUL_values}.

Our new $Q_{\textrm{UL}}$ values definitively show that the potential UL Q value decay for $^{112}$Ag identified in Ref. \cite{gam2019} is not energetically allowed, and that the potential UL Q value decay for $^{115}$Cd identified in Ref. \cite{Haa2013} is not $<$ 1 keV and hence not ultra-low. The fact that our new $Q_{gs}$ value for $^{113}$Ag deviates from the AME value by 69 keV, means that the potential UL Q value decay to the 1/2$^{+}$ state in $^{113}$Cd at 2015.6 keV is $\gg$ 1 keV and is ruled out. However, there is another 1/2$^{+}$ state in $^{113}$Cd at 2080(10) keV that an UL Q value decay could potentially go to. Currently, the 10 keV uncertainty in the energy of this state is too large to say anything more definitive.

\begin{table}[ht]
 \caption{\label{table:QUL_values} Q values for potential UL decay branches identified in Refs. \cite{Haa2013,gam2019}. The $Q_{\textrm{UL}}$ values were obtained using the gs--gs Q values, $Q_\mathrm{CPT}$ and $Q_\mathrm{AME}$, from Table \ref{table:Q}, and excited state energies, $E^{*}$ from Ref. \cite{nndc}. All values listed are in keV.}
\begin{ruledtabular}
\begin{tabular}{cccc}
\multirow{2}{*}{Decay} & \multirow{2}{*}{$E^{*}$} & \multicolumn{2}{c}{$Q_{\textrm{UL}}$}\\
 & & CPT & AME\\
\hline
$^{112}$Ag $\rightarrow$ $^{112}$Cd & 3997.75(14) & -7.59(26) & -6.6(2.4) \\
\hline
\multirow{2}{*}{$^{113}$Ag $\rightarrow$ $^{113}$Cd} & \multicolumn{1}{c}{2015.6(2.5)} & \multicolumn{1}{c}{70.1(5.2)} &
\multicolumn{1}{c}{0.9(16.8)}
\\
& \multicolumn{1}{c}{2080(10)} & \multicolumn{1}{c}{5.7(11.0)} & \multicolumn{1}{c}{-63.5(19.4)} \\
\hline
$^{115}$Cd $\rightarrow$ $^{115}$In & 1448.787(9) & 2.57(34) & 3.1(0.7) \\
\end{tabular}
\end{ruledtabular}
\end{table}

\subsection{Mass Excesses for $^{112,113}$Ag and $^{115}$Cd}\label{SubSection_MEs}

The ratios in Table~\ref{table:Ratios} were used to obtain absolute atomic masses for the parent nuclides, $^{112,113}$Ag, and $^{115}$Cd, via
\begin{equation}\label{Eqn:Mass}
M_p = (M_d - m_e)\bar{R}^{-1} + m_e, 
\end{equation}
with corresponding values for $M_d$ taken from Ref. \cite{Wang2021}. Mass excesses were then obtained and are listed in Table~\ref{table:mass} where they are compared with the values from the AME2020~\cite{Wang2021}. 

\begin{table}[b]
\caption{\label{table:mass} Mass excesses for $^{112}$Ag, $^{113}$Ag and $^{115}$Cd obtained in this work along with results from the AME2020~\cite{Wang2021}\ and the difference $\Delta$ME = ME$_{\rm{CPT}}$ -- ME$_{\rm{AME}}$}
\begin{ruledtabular}
\begin{tabular}{cccc}
\multirow{2}{*}{Isotope} & \multicolumn{1}{c}{This work} & \multicolumn{1}{c}{AME2020} & \multicolumn{1}{c}{$\Delta$ME}\\
 & \multicolumn{1}{c}{(keV/$c^2$)} & \multicolumn{1}{c}{(keV/$c^2$)} & \multicolumn{1}{c}{(keV/$c^2$)}\\
\hline
$^{112}$Ag & $-86\ 584.70(33)$ & $-86\ 583.7(2.4)$ & -1.0(2.4)\\
\hline
$^{113}$Ag & $-86\ 957.6(4.6)$ & $-87\ 026.8(16.6)$ & 69.2(17.2)\\
\hline
$^{115}$Cd & $-88\ 085.00(42)$ & $-88\ 084.5(0.7)$ & -0.5(0.8)\\
\end{tabular}
\end{ruledtabular}
\end{table} 

The mass excesses for $^{112}$Ag and $^{115}$Cd are in good agreement with the AME2020 data, but are factors of approximately seven and two more precise, respectively. Our result for $^{113}$Ag shows that it is less bound by 69.2(17.2) keV compared to the AME2020 value, a 4$\sigma$ discrepancy. The mass of $^{113}$Ag in AME2020 is derived almost entirely from three $^{113}$Ag $\beta$-decay measurements~\cite{Matumoto1970,Fogelberg1990}, the most precise of which is listed as a private communication to the Nuclear Data Group from 1957~\cite{Wang2021}. The $^{115}$Cd mass, on the other hand is derived from a $(d,p)$ reaction linking it to $^{114}$Cd, and the $^{112}$Ag mass is from a Penning trap measurement performed by the ISOLTRAP group \cite{Breitenfeldt2010}. We note that we did observe the $^{113m}$Ag isomer in addition to the $^{113}$Ag ground state, as shown in Fig. \ref{Fig:113Ag_PI-ICR}, and determined their mass difference to be 40.4(4.6) keV/$c^2$, consistent with the literature value of 43.5(1) keV~\cite{Kondev2021} for the energy of the isomeric state.

\section{Conclusion}

We have performed precise determinations of the gs--gs $\beta$-decay Q values of $^{112,113}$Ag, and $^{115}$Cd by measuring the cyclotron frequency ratios of singly-charged parent and daughter ions with the Canadian Penning Trap mass spectrometer. By comparing these Q values with excited state energy levels in the daughter nuclei, the $Q_{\textrm{UL}}$ values for potential UL Q value decays of $^{112}$Ag and $^{115}$Cd  were found to be --7.59(26) keV and 2.57(34) keV respectively. The former is not energetically allowed and the latter is too large (i.e. $>$ 1 keV) to be considered as an UL decay, ruling both out as potential UL Q value $\beta$-decays. Our Q value measurement of $^{113}$Ag indicated a 69(17) keV discrepancy compared to data from the AME2020. This result ruled out the potential UL Q value decay to the 1/2$^{+}$, 2015.6 keV state in $^{113}$Cd, but indicated a new potential UL Q value branch to the 1/2$^{+}$ at 2080(10) keV. Hence, $^{113}$Ag is still a potential candidate with a $Q_{\textrm{UL}}$ of 6(11) keV, but a more precise determination of the $^{113}$Cd(1/2$^{+}$, 2080 keV) state energy will be required to further evaluate this decay branch.

We also report improved atomic masses for  $^{112,113}$Ag and $^{115}$Cd. The atomic masses of $^{112}$Ag and $^{115}$Cd are in good agreement with the AME2020 values, and have had their uncertainties reduced by factors of seven and two, respectively. We observed an $\approx$70 keV discrepancy in the mass of $^{113}$Ag compared to the AME.

\section*{Acknowledgments}

This material is based upon work supported by the US Department of Energy, Office of Science, Office of Nuclear Physics under contract No. DE-AC02-06CH11357 and Award No. DE-SC0015927, by NSERC (Canada), Application No. SAPPJ-2015-00034 and SAPPJ-2018-00028, and by the National Science Foundation under grant No. PHY-1713857. Support was also provided by Central Michigan University.

\bibliography{ULQV_paper.bib, MR_Refs.bib, ref.bib} 

\begin{thebibliography}{48}%
\makeatletter
\providecommand \@ifxundefined [1]{%
 \@ifx{#1\undefined}
}%
\providecommand \@ifnum [1]{%
 \ifnum #1\expandafter \@firstoftwo
 \else \expandafter \@secondoftwo
 \fi
}%
\providecommand \@ifx [1]{%
 \ifx #1\expandafter \@firstoftwo
 \else \expandafter \@secondoftwo
 \fi
}%
\providecommand \natexlab [1]{#1}%
\providecommand \enquote  [1]{``#1''}%
\providecommand \bibnamefont  [1]{#1}%
\providecommand \bibfnamefont [1]{#1}%
\providecommand \citenamefont [1]{#1}%
\providecommand \href@noop [0]{\@secondoftwo}%
\providecommand \href [0]{\begingroup \@sanitize@url \@href}%
\providecommand \@href[1]{\@@startlink{#1}\@@href}%
\providecommand \@@href[1]{\endgroup#1\@@endlink}%
\providecommand \@sanitize@url [0]{\catcode `\\12\catcode `\$12\catcode
  `\&12\catcode `\#12\catcode `\^12\catcode `\_12\catcode `\%12\relax}%
\providecommand \@@startlink[1]{}%
\providecommand \@@endlink[0]{}%
\providecommand \url  [0]{\begingroup\@sanitize@url \@url }%
\providecommand \@url [1]{\endgroup\@href {#1}{\urlprefix }}%
\providecommand \urlprefix  [0]{URL }%
\providecommand \Eprint [0]{\href }%
\providecommand \doibase [0]{http://dx.doi.org/}%
\providecommand \selectlanguage [0]{\@gobble}%
\providecommand \bibinfo  [0]{\@secondoftwo}%
\providecommand \bibfield  [0]{\@secondoftwo}%
\providecommand \translation [1]{[#1]}%
\providecommand \BibitemOpen [0]{}%
\providecommand \bibitemStop [0]{}%
\providecommand \bibitemNoStop [0]{.\EOS\space}%
\providecommand \EOS [0]{\spacefactor3000\relax}%
\providecommand \BibitemShut  [1]{\csname bibitem#1\endcsname}%
\let\auto@bib@innerbib\@empty
\bibitem [{\citenamefont {Severijns}\ \emph {et~al.}(2006)\citenamefont
  {Severijns}, \citenamefont {Beck},\ and\ \citenamefont
  {Naviliat-Cuncic}}]{Severijns2006}%
  \BibitemOpen
  \bibfield  {author} {\bibinfo {author} {\bibfnamefont {N.}~\bibnamefont
  {Severijns}}, \bibinfo {author} {\bibfnamefont {M.}~\bibnamefont {Beck}}, \
  and\ \bibinfo {author} {\bibfnamefont {O.}~\bibnamefont {Naviliat-Cuncic}},\
  }\href {\doibase 10.1103/RevModPhys.78.991} {\bibfield  {journal} {\bibinfo
  {journal} {Rev. Mod. Phys.}\ }\textbf {\bibinfo {volume} {78}},\ \bibinfo
  {pages} {991} (\bibinfo {year} {2006})}\BibitemShut {NoStop}%
\bibitem [{\citenamefont {Haaranen}\ \emph {et~al.}(2016)\citenamefont
  {Haaranen}, \citenamefont {Srivastava},\ and\ \citenamefont
  {Suhonen}}]{Haaranen2016_SSM}%
  \BibitemOpen
  \bibfield  {author} {\bibinfo {author} {\bibfnamefont {M.}~\bibnamefont
  {Haaranen}}, \bibinfo {author} {\bibfnamefont {P.~C.}\ \bibnamefont
  {Srivastava}}, \ and\ \bibinfo {author} {\bibfnamefont {J.}~\bibnamefont
  {Suhonen}},\ }\href {\doibase 10.1103/PhysRevC.93.034308} {\bibfield
  {journal} {\bibinfo  {journal} {Phys. Rev. C}\ }\textbf {\bibinfo {volume}
  {93}},\ \bibinfo {pages} {034308} (\bibinfo {year} {2016})}\BibitemShut
  {NoStop}%
\bibitem [{\citenamefont {Aker}\ \emph {et~al.}(2019)\citenamefont {Aker},
  \citenamefont {Altenm\"uller}, \citenamefont {Arenz}, \citenamefont
  {Babutzka}, \citenamefont {Barrett}, \citenamefont {Bauer}, \citenamefont
  {Beck}, \citenamefont {Beglarian}, \citenamefont {Behrens}, \citenamefont
  {Bergmann} \emph {et~al.}}]{Aker2021}%
  \BibitemOpen
  \bibfield  {author} {\bibinfo {author} {\bibfnamefont {M.}~\bibnamefont
  {Aker}}, \bibinfo {author} {\bibfnamefont {K.}~\bibnamefont {Altenm\"uller}},
  \bibinfo {author} {\bibfnamefont {M.}~\bibnamefont {Arenz}}, \bibinfo
  {author} {\bibfnamefont {M.}~\bibnamefont {Babutzka}}, \bibinfo {author}
  {\bibfnamefont {J.}~\bibnamefont {Barrett}}, \bibinfo {author} {\bibfnamefont
  {S.}~\bibnamefont {Bauer}}, \bibinfo {author} {\bibfnamefont
  {M.}~\bibnamefont {Beck}}, \bibinfo {author} {\bibfnamefont {A.}~\bibnamefont
  {Beglarian}}, \bibinfo {author} {\bibfnamefont {J.}~\bibnamefont {Behrens}},
  \bibinfo {author} {\bibfnamefont {T.}~\bibnamefont {Bergmann}},  \emph
  {et~al.} (\bibinfo {collaboration} {KATRIN Collaboration}),\ }\href {\doibase
  10.1103/PhysRevLett.123.221802} {\bibfield  {journal} {\bibinfo  {journal}
  {Phys. Rev. Lett.}\ }\textbf {\bibinfo {volume} {123}},\ \bibinfo {pages}
  {221802} (\bibinfo {year} {2019})}\BibitemShut {NoStop}%
\bibitem [{\citenamefont {Gastaldo}\ \emph {et~al.}(2017)\citenamefont
  {Gastaldo} \emph {et~al.}}]{Gas2017}%
  \BibitemOpen
  \bibfield  {author} {\bibinfo {author} {\bibfnamefont {L.}~\bibnamefont
  {Gastaldo}} \emph {et~al.} (\bibinfo {collaboration} {ECHo Collaboration}),\
  }\href {\doibase 10.1140/epjst/e2017-70071-y} {\bibfield  {journal} {\bibinfo
   {journal} {The European Physical Journal Special Topics}\ }\textbf {\bibinfo
  {volume} {226}},\ \bibinfo {pages} {1623} (\bibinfo {year}
  {2017})}\BibitemShut {NoStop}%
\bibitem [{\citenamefont {Alpert}\ \emph {et~al.}(2015)\citenamefont {Alpert}
  \emph {et~al.}}]{Alp2015}%
  \BibitemOpen
  \bibfield  {author} {\bibinfo {author} {\bibfnamefont {B.}~\bibnamefont
  {Alpert}} \emph {et~al.} (\bibinfo {collaboration} {HOLMES Collaboration}),\
  }\href {\doibase 10.1140/epjc/s10052-015-3329-5} {\bibfield  {journal}
  {\bibinfo  {journal} {The European Physical Journal C}\ }\textbf {\bibinfo
  {volume} {75}},\ \bibinfo {pages} {112} (\bibinfo {year} {2015})}\BibitemShut
  {NoStop}%
\bibitem [{\citenamefont {Arnaboldi}\ \emph {et~al.}(2003)\citenamefont
  {Arnaboldi}, \citenamefont {Brofferio}, \citenamefont {Cremonesi},
  \citenamefont {Fiorini}, \citenamefont {Lo~Bianco}, \citenamefont
  {Martensson}, \citenamefont {Nucciotti}, \citenamefont {Pavan}, \citenamefont
  {Pessina}, \citenamefont {Pirro}, \citenamefont {Previtali}, \citenamefont
  {Sisti}, \citenamefont {Giuliani}, \citenamefont {Margesin},\ and\
  \citenamefont {Zen}}]{Arnaboldi2003}%
  \BibitemOpen
  \bibfield  {author} {\bibinfo {author} {\bibfnamefont {C.}~\bibnamefont
  {Arnaboldi}}, \bibinfo {author} {\bibfnamefont {C.}~\bibnamefont
  {Brofferio}}, \bibinfo {author} {\bibfnamefont {O.}~\bibnamefont
  {Cremonesi}}, \bibinfo {author} {\bibfnamefont {E.}~\bibnamefont {Fiorini}},
  \bibinfo {author} {\bibfnamefont {C.}~\bibnamefont {Lo~Bianco}}, \bibinfo
  {author} {\bibfnamefont {L.}~\bibnamefont {Martensson}}, \bibinfo {author}
  {\bibfnamefont {A.}~\bibnamefont {Nucciotti}}, \bibinfo {author}
  {\bibfnamefont {M.}~\bibnamefont {Pavan}}, \bibinfo {author} {\bibfnamefont
  {G.}~\bibnamefont {Pessina}}, \bibinfo {author} {\bibfnamefont
  {S.}~\bibnamefont {Pirro}}, \bibinfo {author} {\bibfnamefont
  {E.}~\bibnamefont {Previtali}}, \bibinfo {author} {\bibfnamefont
  {M.}~\bibnamefont {Sisti}}, \bibinfo {author} {\bibfnamefont
  {A.}~\bibnamefont {Giuliani}}, \bibinfo {author} {\bibfnamefont
  {B.}~\bibnamefont {Margesin}}, \ and\ \bibinfo {author} {\bibfnamefont
  {M.}~\bibnamefont {Zen}},\ }\href {\doibase 10.1103/PhysRevLett.91.161802}
  {\bibfield  {journal} {\bibinfo  {journal} {Phys. Rev. Lett.}\ }\textbf
  {\bibinfo {volume} {91}},\ \bibinfo {pages} {161802} (\bibinfo {year}
  {2003})}\BibitemShut {NoStop}%
\bibitem [{\citenamefont {Chechev}(2011)}]{Chechev2011}%
  \BibitemOpen
  \bibfield  {author} {\bibinfo {author} {\bibfnamefont {V.~P.}\ \bibnamefont
  {Chechev}},\ }\href {\doibase 10.1134/S106377881111007X} {\bibfield
  {journal} {\bibinfo  {journal} {Phy. At. Nucl.}\ }\textbf {\bibinfo {volume}
  {74}},\ \bibinfo {pages} {1713} (\bibinfo {year} {2011})}\BibitemShut
  {NoStop}%
\bibitem [{\citenamefont {Kostensalo}\ and\ \citenamefont
  {Suhonen}(2018)}]{Suhonen2018}%
  \BibitemOpen
  \bibfield  {author} {\bibinfo {author} {\bibfnamefont {J.}~\bibnamefont
  {Kostensalo}}\ and\ \bibinfo {author} {\bibfnamefont {J.}~\bibnamefont
  {Suhonen}},\ }\href {\doibase 10.1142/S0217751X1843008X} {\bibfield
  {journal} {\bibinfo  {journal} {Int. J. Mod. Phys. A}\ }\textbf {\bibinfo
  {volume} {33}},\ \bibinfo {pages} {1843008} (\bibinfo {year}
  {2018})}\BibitemShut {NoStop}%
\bibitem [{\citenamefont {Mustonen}\ and\ \citenamefont
  {Suhonen}(2010{\natexlab{a}})}]{Mustonen2010}%
  \BibitemOpen
  \bibfield  {author} {\bibinfo {author} {\bibfnamefont {M.~T.}\ \bibnamefont
  {Mustonen}}\ and\ \bibinfo {author} {\bibfnamefont {J.}~\bibnamefont
  {Suhonen}},\ }\href {\doibase https://doi.org/10.1063/1.3527233} {\bibfield
  {journal} {\bibinfo  {journal} {AIP Conference Proceedings}\ }\textbf
  {\bibinfo {volume} {1304}},\ \bibinfo {pages} {401} (\bibinfo {year}
  {2010}{\natexlab{a}})}\BibitemShut {NoStop}%
\bibitem [{\citenamefont {Mustonen}\ and\ \citenamefont
  {Suhonen}(2010{\natexlab{b}})}]{mus2010}%
  \BibitemOpen
  \bibfield  {author} {\bibinfo {author} {\bibfnamefont {M.~T.}\ \bibnamefont
  {Mustonen}}\ and\ \bibinfo {author} {\bibfnamefont {J.}~\bibnamefont
  {Suhonen}},\ }\href {\doibase 10.1088/0954-3899/37/6/064008} {\bibfield
  {journal} {\bibinfo  {journal} {J. Phys. G: Nucl. Part. Phys.}\ }\textbf
  {\bibinfo {volume} {37}},\ \bibinfo {pages} {064008} (\bibinfo {year}
  {2010}{\natexlab{b}})}\BibitemShut {NoStop}%
\bibitem [{\citenamefont {Cattadori}\ \emph {et~al.}(2004)\citenamefont
  {Cattadori}, \citenamefont {Deo}, \citenamefont {Laubenstein}, \citenamefont
  {Pandola},\ and\ \citenamefont {Tretyak}}]{cat2005}%
  \BibitemOpen
  \bibfield  {author} {\bibinfo {author} {\bibfnamefont {C.~M.}\ \bibnamefont
  {Cattadori}}, \bibinfo {author} {\bibfnamefont {M.~D.}\ \bibnamefont {Deo}},
  \bibinfo {author} {\bibfnamefont {M.}~\bibnamefont {Laubenstein}}, \bibinfo
  {author} {\bibfnamefont {L.}~\bibnamefont {Pandola}}, \ and\ \bibinfo
  {author} {\bibfnamefont {V.~I.}\ \bibnamefont {Tretyak}},\ }\href {\doibase
  10.1016/j.nuclphysa.2004.10.025} {\bibfield  {journal} {\bibinfo  {journal}
  {Nucl. Phys. A}\ }\textbf {\bibinfo {volume} {748}},\ \bibinfo {pages} {333}
  (\bibinfo {year} {2004})}\BibitemShut {NoStop}%
\bibitem [{\citenamefont {Wieslander}\ \emph
  {et~al.}(2009{\natexlab{a}})\citenamefont {Wieslander}, \citenamefont
  {Suhonen}, \citenamefont {Eronen}, \citenamefont {Hult}, \citenamefont
  {Elomaa}, \citenamefont {Jokinen}, \citenamefont {Marissens}, \citenamefont
  {Misiaszek}, \citenamefont {Mustonen}, \citenamefont {Rahaman}, \citenamefont
  {Weber},\ and\ \citenamefont {\"{A}yst\"{o}}}]{wie2009}%
  \BibitemOpen
  \bibfield  {author} {\bibinfo {author} {\bibfnamefont {J.~S.~E.}\
  \bibnamefont {Wieslander}}, \bibinfo {author} {\bibfnamefont
  {J.}~\bibnamefont {Suhonen}}, \bibinfo {author} {\bibfnamefont
  {T.}~\bibnamefont {Eronen}}, \bibinfo {author} {\bibfnamefont
  {M.}~\bibnamefont {Hult}}, \bibinfo {author} {\bibfnamefont {V.-V.}\
  \bibnamefont {Elomaa}}, \bibinfo {author} {\bibfnamefont {A.}~\bibnamefont
  {Jokinen}}, \bibinfo {author} {\bibfnamefont {G.}~\bibnamefont {Marissens}},
  \bibinfo {author} {\bibfnamefont {M.}~\bibnamefont {Misiaszek}}, \bibinfo
  {author} {\bibfnamefont {M.~T.}\ \bibnamefont {Mustonen}}, \bibinfo {author}
  {\bibfnamefont {S.}~\bibnamefont {Rahaman}}, \bibinfo {author} {\bibfnamefont
  {C.}~\bibnamefont {Weber}}, \ and\ \bibinfo {author} {\bibfnamefont
  {J.}~\bibnamefont {\"{A}yst\"{o}}},\ }\href
  {https://doi.org/10.1103/PhysRevLett.103.122501} {\bibfield  {journal}
  {\bibinfo  {journal} {Phys. Rev. Lett.}\ }\textbf {\bibinfo {volume} {103}},\
  \bibinfo {pages} {122501} (\bibinfo {year} {2009}{\natexlab{a}})}\BibitemShut
  {NoStop}%
\bibitem [{\citenamefont {Andreotti}\ \emph {et~al.}(2011)\citenamefont
  {Andreotti}, \citenamefont {Hult}, \citenamefont {Gonz\'alez~de Ordu\~na},
  \citenamefont {Marissens}, \citenamefont {Wieslander},\ and\ \citenamefont
  {Misiaszek}}]{And2011}%
  \BibitemOpen
  \bibfield  {author} {\bibinfo {author} {\bibfnamefont {E.}~\bibnamefont
  {Andreotti}}, \bibinfo {author} {\bibfnamefont {M.}~\bibnamefont {Hult}},
  \bibinfo {author} {\bibfnamefont {R.}~\bibnamefont {Gonz\'alez~de Ordu\~na}},
  \bibinfo {author} {\bibfnamefont {G.}~\bibnamefont {Marissens}}, \bibinfo
  {author} {\bibfnamefont {J.~S.~E.}\ \bibnamefont {Wieslander}}, \ and\
  \bibinfo {author} {\bibfnamefont {M.}~\bibnamefont {Misiaszek}},\ }\href
  {\doibase 10.1103/PhysRevC.84.044605} {\bibfield  {journal} {\bibinfo
  {journal} {Phys. Rev. C}\ }\textbf {\bibinfo {volume} {84}},\ \bibinfo
  {pages} {044605} (\bibinfo {year} {2011})}\BibitemShut {NoStop}%
\bibitem [{\citenamefont {Mount}\ \emph {et~al.}(2009)\citenamefont {Mount},
  \citenamefont {Redshaw},\ and\ \citenamefont {Myers}}]{Mount2009}%
  \BibitemOpen
  \bibfield  {author} {\bibinfo {author} {\bibfnamefont {B.~J.}\ \bibnamefont
  {Mount}}, \bibinfo {author} {\bibfnamefont {M.}~\bibnamefont {Redshaw}}, \
  and\ \bibinfo {author} {\bibfnamefont {E.~G.}\ \bibnamefont {Myers}},\ }\href
  {\doibase 10.1103/PhysRevLett.103.122502} {\bibfield  {journal} {\bibinfo
  {journal} {Phys. Rev. Lett.}\ }\textbf {\bibinfo {volume} {103}},\ \bibinfo
  {pages} {122502} (\bibinfo {year} {2009})}\BibitemShut {NoStop}%
\bibitem [{\citenamefont {Wieslander}\ \emph
  {et~al.}(2009{\natexlab{b}})\citenamefont {Wieslander}, \citenamefont
  {Suhonen}, \citenamefont {Eronen}, \citenamefont {Hult}, \citenamefont
  {Elomaa}, \citenamefont {Jokinen}, \citenamefont {Marissens}, \citenamefont
  {Misiaszek}, \citenamefont {Mustonen}, \citenamefont {Rahaman}, \citenamefont
  {Weber},\ and\ \citenamefont {\"Ayst\"o}}]{Wieslander2009}%
  \BibitemOpen
  \bibfield  {author} {\bibinfo {author} {\bibfnamefont {J.~S.~E.}\
  \bibnamefont {Wieslander}}, \bibinfo {author} {\bibfnamefont
  {J.}~\bibnamefont {Suhonen}}, \bibinfo {author} {\bibfnamefont
  {T.}~\bibnamefont {Eronen}}, \bibinfo {author} {\bibfnamefont
  {M.}~\bibnamefont {Hult}}, \bibinfo {author} {\bibfnamefont {V.-V.}\
  \bibnamefont {Elomaa}}, \bibinfo {author} {\bibfnamefont {A.}~\bibnamefont
  {Jokinen}}, \bibinfo {author} {\bibfnamefont {G.}~\bibnamefont {Marissens}},
  \bibinfo {author} {\bibfnamefont {M.}~\bibnamefont {Misiaszek}}, \bibinfo
  {author} {\bibfnamefont {M.~T.}\ \bibnamefont {Mustonen}}, \bibinfo {author}
  {\bibfnamefont {S.}~\bibnamefont {Rahaman}}, \bibinfo {author} {\bibfnamefont
  {C.}~\bibnamefont {Weber}}, \ and\ \bibinfo {author} {\bibfnamefont
  {J.}~\bibnamefont {\"Ayst\"o}},\ }\href {\doibase
  10.1103/PhysRevLett.103.122501} {\bibfield  {journal} {\bibinfo  {journal}
  {Phys. Rev. Lett.}\ }\textbf {\bibinfo {volume} {103}},\ \bibinfo {pages}
  {122501} (\bibinfo {year} {2009}{\natexlab{b}})}\BibitemShut {NoStop}%
\bibitem [{\citenamefont {Wang}\ \emph {et~al.}(2017)\citenamefont {Wang},
  \citenamefont {Audi}, \citenamefont {Kondev}, \citenamefont {Huang},
  \citenamefont {Naimi},\ and\ \citenamefont {Xu}}]{Wan2017}%
  \BibitemOpen
  \bibfield  {author} {\bibinfo {author} {\bibfnamefont {M.}~\bibnamefont
  {Wang}}, \bibinfo {author} {\bibfnamefont {G.}~\bibnamefont {Audi}}, \bibinfo
  {author} {\bibfnamefont {F.}~\bibnamefont {Kondev}}, \bibinfo {author}
  {\bibfnamefont {W.}~\bibnamefont {Huang}}, \bibinfo {author} {\bibfnamefont
  {S.}~\bibnamefont {Naimi}}, \ and\ \bibinfo {author} {\bibfnamefont
  {X.}~\bibnamefont {Xu}},\ }\href {\doibase 10.1088/1674-1137/41/3/030002}
  {\bibfield  {journal} {\bibinfo  {journal} {Chin. Phys. C}\ }\textbf
  {\bibinfo {volume} {41}},\ \bibinfo {pages} {030003} (\bibinfo {year}
  {2017})}\BibitemShut {NoStop}%
\bibitem [{\citenamefont {Zheltonozhsky}\ \emph {et~al.}(2018)\citenamefont
  {Zheltonozhsky}, \citenamefont {Savrasov}, \citenamefont {Strilchuk},\ and\
  \citenamefont {Tretyak}}]{Zhe2019}%
  \BibitemOpen
  \bibfield  {author} {\bibinfo {author} {\bibfnamefont {V.~A.}\ \bibnamefont
  {Zheltonozhsky}}, \bibinfo {author} {\bibfnamefont {A.~M.}\ \bibnamefont
  {Savrasov}}, \bibinfo {author} {\bibfnamefont {N.~V.}\ \bibnamefont
  {Strilchuk}}, \ and\ \bibinfo {author} {\bibfnamefont {V.~I.}\ \bibnamefont
  {Tretyak}},\ }\href {\doibase 10.1209/0295-5075/121/12001} {\bibfield
  {journal} {\bibinfo  {journal} {Europhys. Lett.}\ }\textbf {\bibinfo {volume}
  {121}},\ \bibinfo {pages} {12001} (\bibinfo {year} {2018})}\BibitemShut
  {NoStop}%
\bibitem [{\citenamefont {Suhonen}\ and\ \citenamefont
  {Mustonen}(2010)}]{Suh2010}%
  \BibitemOpen
  \bibfield  {author} {\bibinfo {author} {\bibfnamefont {J.}~\bibnamefont
  {Suhonen}}\ and\ \bibinfo {author} {\bibfnamefont {M.}~\bibnamefont
  {Mustonen}},\ }\href {\doibase https://doi.org/10.1016/j.ppnp.2009.12.018}
  {\bibfield  {journal} {\bibinfo  {journal} {Progress in Particle and Nuclear
  Physics}\ }\textbf {\bibinfo {volume} {64}},\ \bibinfo {pages} {235 }
  (\bibinfo {year} {2010})},\ \bibinfo {note} {neutrinos in Cosmology, in
  Astro, Particle and Nuclear Physics}\BibitemShut {NoStop}%
\bibitem [{\citenamefont {Ferri}\ \emph {et~al.}(2015)\citenamefont {Ferri},
  \citenamefont {Bagliani}, \citenamefont {Biasotti}, \citenamefont {Ceruti},
  \citenamefont {Corsini}, \citenamefont {Faverzani}, \citenamefont {Gatti},
  \citenamefont {Giachero}, \citenamefont {Gotti}, \citenamefont {Kilbourne}
  \emph {et~al.}}]{Ferri2015}%
  \BibitemOpen
  \bibfield  {author} {\bibinfo {author} {\bibfnamefont {E.}~\bibnamefont
  {Ferri}}, \bibinfo {author} {\bibfnamefont {D.}~\bibnamefont {Bagliani}},
  \bibinfo {author} {\bibfnamefont {M.}~\bibnamefont {Biasotti}}, \bibinfo
  {author} {\bibfnamefont {G.}~\bibnamefont {Ceruti}}, \bibinfo {author}
  {\bibfnamefont {D.}~\bibnamefont {Corsini}}, \bibinfo {author} {\bibfnamefont
  {M.}~\bibnamefont {Faverzani}}, \bibinfo {author} {\bibfnamefont
  {F.}~\bibnamefont {Gatti}}, \bibinfo {author} {\bibfnamefont
  {A.}~\bibnamefont {Giachero}}, \bibinfo {author} {\bibfnamefont
  {C.}~\bibnamefont {Gotti}}, \bibinfo {author} {\bibfnamefont
  {C.}~\bibnamefont {Kilbourne}},  \emph {et~al.},\ }\href@noop {} {\bibfield
  {journal} {\bibinfo  {journal} {Phys. Procedia}\ }\textbf {\bibinfo {volume}
  {61}},\ \bibinfo {pages} {227} (\bibinfo {year} {2015})}\BibitemShut
  {NoStop}%
\bibitem [{\citenamefont {Haaranen}\ and\ \citenamefont
  {Suhonen}(2013)}]{Haa2013}%
  \BibitemOpen
  \bibfield  {author} {\bibinfo {author} {\bibfnamefont {M.}~\bibnamefont
  {Haaranen}}\ and\ \bibinfo {author} {\bibfnamefont {J.}~\bibnamefont
  {Suhonen}},\ }\href {\doibase 10.1140/epja/i2013-13093-8} {\bibfield
  {journal} {\bibinfo  {journal} {The European Physical Journal A}\ }\textbf
  {\bibinfo {volume} {49}},\ \bibinfo {pages} {93} (\bibinfo {year}
  {2013})}\BibitemShut {NoStop}%
\bibitem [{\citenamefont {Mustonen}\ and\ \citenamefont
  {Suhonen}(2011)}]{Mus2011}%
  \BibitemOpen
  \bibfield  {author} {\bibinfo {author} {\bibfnamefont {M.}~\bibnamefont
  {Mustonen}}\ and\ \bibinfo {author} {\bibfnamefont {J.}~\bibnamefont
  {Suhonen}},\ }\href {\doibase https://doi.org/10.1016/j.physletb.2011.07.088}
  {\bibfield  {journal} {\bibinfo  {journal} {Physics Letters B}\ }\textbf
  {\bibinfo {volume} {703}},\ \bibinfo {pages} {370 } (\bibinfo {year}
  {2011})}\BibitemShut {NoStop}%
\bibitem [{\citenamefont {Wang}\ \emph {et~al.}(2021)\citenamefont {Wang},
  \citenamefont {Huang}, \citenamefont {Kondev}, \citenamefont {Audi},\ and\
  \citenamefont {Naimi}}]{Wang2021}%
  \BibitemOpen
  \bibfield  {author} {\bibinfo {author} {\bibfnamefont {M.}~\bibnamefont
  {Wang}}, \bibinfo {author} {\bibfnamefont {W.}~\bibnamefont {Huang}},
  \bibinfo {author} {\bibfnamefont {F.}~\bibnamefont {Kondev}}, \bibinfo
  {author} {\bibfnamefont {G.}~\bibnamefont {Audi}}, \ and\ \bibinfo {author}
  {\bibfnamefont {S.}~\bibnamefont {Naimi}},\ }\href {\doibase
  10.1088/1674-1137/abddaf} {\bibfield  {journal} {\bibinfo  {journal} {Chinese
  Physics C}\ }\textbf {\bibinfo {volume} {45}},\ \bibinfo {pages} {030003}
  (\bibinfo {year} {2021})}\BibitemShut {NoStop}%
\bibitem [{\citenamefont {de~Roubin}\ \emph {et~al.}(2020)\citenamefont
  {de~Roubin}, \citenamefont {Kostensalo}, \citenamefont {Eronen},
  \citenamefont {Canete}, \citenamefont {de~Groote}, \citenamefont {Jokinen},
  \citenamefont {Kankainen}, \citenamefont {Nesterenko}, \citenamefont {Moore},
  \citenamefont {Rinta-Antila}, \citenamefont {Suhonen},\ and\ \citenamefont
  {Vil\'en}}]{deRoubin2020}%
  \BibitemOpen
  \bibfield  {author} {\bibinfo {author} {\bibfnamefont {A.}~\bibnamefont
  {de~Roubin}}, \bibinfo {author} {\bibfnamefont {J.}~\bibnamefont
  {Kostensalo}}, \bibinfo {author} {\bibfnamefont {T.}~\bibnamefont {Eronen}},
  \bibinfo {author} {\bibfnamefont {L.}~\bibnamefont {Canete}}, \bibinfo
  {author} {\bibfnamefont {R.~P.}\ \bibnamefont {de~Groote}}, \bibinfo {author}
  {\bibfnamefont {A.}~\bibnamefont {Jokinen}}, \bibinfo {author} {\bibfnamefont
  {A.}~\bibnamefont {Kankainen}}, \bibinfo {author} {\bibfnamefont {D.~A.}\
  \bibnamefont {Nesterenko}}, \bibinfo {author} {\bibfnamefont {I.~D.}\
  \bibnamefont {Moore}}, \bibinfo {author} {\bibfnamefont {S.}~\bibnamefont
  {Rinta-Antila}}, \bibinfo {author} {\bibfnamefont {J.}~\bibnamefont
  {Suhonen}}, \ and\ \bibinfo {author} {\bibfnamefont {M.}~\bibnamefont
  {Vil\'en}},\ }\href {\doibase 10.1103/PhysRevLett.124.222503} {\bibfield
  {journal} {\bibinfo  {journal} {Phys. Rev. Lett.}\ }\textbf {\bibinfo
  {volume} {124}},\ \bibinfo {pages} {222503} (\bibinfo {year}
  {2020})}\BibitemShut {NoStop}%
\bibitem [{\citenamefont {Mustonen}\ and\ \citenamefont
  {Suhonen}(2010{\natexlab{c}})}]{mus2010_2}%
  \BibitemOpen
  \bibfield  {author} {\bibinfo {author} {\bibfnamefont {M.~T.}\ \bibnamefont
  {Mustonen}}\ and\ \bibinfo {author} {\bibfnamefont {J.}~\bibnamefont
  {Suhonen}},\ }\href {https://doi.org/10.1063/1.3527233} {\bibfield  {journal}
  {\bibinfo  {journal} {AIP Conf. Proc.}\ }\textbf {\bibinfo {volume} {1304}},\
  \bibinfo {pages} {401} (\bibinfo {year} {2010}{\natexlab{c}})}\BibitemShut
  {NoStop}%
\bibitem [{\citenamefont {Suhonen}(2014)}]{Suh2014}%
  \BibitemOpen
  \bibfield  {author} {\bibinfo {author} {\bibfnamefont {J.}~\bibnamefont
  {Suhonen}},\ }\href {\doibase 10.1088/0031-8949/89/5/054032} {\bibfield
  {journal} {\bibinfo  {journal} {Physica Scripta}\ }\textbf {\bibinfo {volume}
  {89}},\ \bibinfo {pages} {054032} (\bibinfo {year} {2014})}\BibitemShut
  {NoStop}%
\bibitem [{\citenamefont {Kopp}\ and\ \citenamefont {Merle}(2010)}]{kop2010}%
  \BibitemOpen
  \bibfield  {author} {\bibinfo {author} {\bibfnamefont {J.}~\bibnamefont
  {Kopp}}\ and\ \bibinfo {author} {\bibfnamefont {A.}~\bibnamefont {Merle}},\
  }\href {https://doi.org/10.1103/PhysRevC.81.045501} {\bibfield  {journal}
  {\bibinfo  {journal} {Phys. Rev. C}\ }\textbf {\bibinfo {volume} {81}},\
  \bibinfo {pages} {045501} (\bibinfo {year} {2010})}\BibitemShut {NoStop}%
\bibitem [{\citenamefont {Gamage}\ \emph {et~al.}(2019)\citenamefont {Gamage},
  \citenamefont {Bhandari}, \citenamefont {Gamage}, \citenamefont {Sandler},\
  and\ \citenamefont {Redshaw}}]{gam2019}%
  \BibitemOpen
  \bibfield  {author} {\bibinfo {author} {\bibfnamefont {N.~D.}\ \bibnamefont
  {Gamage}}, \bibinfo {author} {\bibfnamefont {R.}~\bibnamefont {Bhandari}},
  \bibinfo {author} {\bibfnamefont {M.~H.}\ \bibnamefont {Gamage}}, \bibinfo
  {author} {\bibfnamefont {R.}~\bibnamefont {Sandler}}, \ and\ \bibinfo
  {author} {\bibfnamefont {M.}~\bibnamefont {Redshaw}},\ }\href
  {hhttps://doi.org/10.1007/s10751-019-1588-51} {\bibfield  {journal} {\bibinfo
   {journal} {Hyp. Int.}\ }\textbf {\bibinfo {volume} {240}},\ \bibinfo {pages}
  {43} (\bibinfo {year} {2019})}\BibitemShut {NoStop}%
\bibitem [{\citenamefont {Sandler}\ \emph {et~al.}(2019)\citenamefont
  {Sandler}, \citenamefont {Bollen}, \citenamefont {Gamage}, \citenamefont
  {Hamaker}, \citenamefont {Izzo}, \citenamefont {Puentes}, \citenamefont
  {Redshaw}, \citenamefont {Ringle},\ and\ \citenamefont
  {Yandow}}]{Sandler2019_89Y}%
  \BibitemOpen
  \bibfield  {author} {\bibinfo {author} {\bibfnamefont {R.}~\bibnamefont
  {Sandler}}, \bibinfo {author} {\bibfnamefont {G.}~\bibnamefont {Bollen}},
  \bibinfo {author} {\bibfnamefont {N.~D.}\ \bibnamefont {Gamage}}, \bibinfo
  {author} {\bibfnamefont {A.}~\bibnamefont {Hamaker}}, \bibinfo {author}
  {\bibfnamefont {C.}~\bibnamefont {Izzo}}, \bibinfo {author} {\bibfnamefont
  {D.}~\bibnamefont {Puentes}}, \bibinfo {author} {\bibfnamefont
  {M.}~\bibnamefont {Redshaw}}, \bibinfo {author} {\bibfnamefont
  {R.}~\bibnamefont {Ringle}}, \ and\ \bibinfo {author} {\bibfnamefont
  {I.}~\bibnamefont {Yandow}},\ }\href {\doibase 10.1103/PhysRevC.100.024309}
  {\bibfield  {journal} {\bibinfo  {journal} {Phys. Rev. C}\ }\textbf {\bibinfo
  {volume} {100}},\ \bibinfo {pages} {024309} (\bibinfo {year}
  {2019})}\BibitemShut {NoStop}%
\bibitem [{\citenamefont {Ge}\ \emph {et~al.}(2021{\natexlab{a}})\citenamefont
  {Ge}, \citenamefont {Eronen}, \citenamefont {de~Roubin}, \citenamefont
  {Nesterenko}, \citenamefont {Hukkanen}, \citenamefont {Beliuskina},
  \citenamefont {de~Groote}, \citenamefont {Geldhof}, \citenamefont {Gins},
  \citenamefont {Kankainen}, \citenamefont {Koszor\'us}, \citenamefont
  {Kotila}, \citenamefont {Kostensalo}, \citenamefont {Moore}, \citenamefont
  {Raggio}, \citenamefont {Rinta-Antila}, \citenamefont {Suhonen},
  \citenamefont {Virtanen}, \citenamefont {Weaver}, \citenamefont
  {Zadvornaya},\ and\ \citenamefont {Jokinen}}]{Ge2021}%
  \BibitemOpen
  \bibfield  {author} {\bibinfo {author} {\bibfnamefont {Z.}~\bibnamefont
  {Ge}}, \bibinfo {author} {\bibfnamefont {T.}~\bibnamefont {Eronen}}, \bibinfo
  {author} {\bibfnamefont {A.}~\bibnamefont {de~Roubin}}, \bibinfo {author}
  {\bibfnamefont {D.~A.}\ \bibnamefont {Nesterenko}}, \bibinfo {author}
  {\bibfnamefont {M.}~\bibnamefont {Hukkanen}}, \bibinfo {author}
  {\bibfnamefont {O.}~\bibnamefont {Beliuskina}}, \bibinfo {author}
  {\bibfnamefont {R.}~\bibnamefont {de~Groote}}, \bibinfo {author}
  {\bibfnamefont {S.}~\bibnamefont {Geldhof}}, \bibinfo {author} {\bibfnamefont
  {W.}~\bibnamefont {Gins}}, \bibinfo {author} {\bibfnamefont {A.}~\bibnamefont
  {Kankainen}}, \bibinfo {author} {\bibfnamefont {A.}~\bibnamefont
  {Koszor\'us}}, \bibinfo {author} {\bibfnamefont {J.}~\bibnamefont {Kotila}},
  \bibinfo {author} {\bibfnamefont {J.}~\bibnamefont {Kostensalo}}, \bibinfo
  {author} {\bibfnamefont {I.~D.}\ \bibnamefont {Moore}}, \bibinfo {author}
  {\bibfnamefont {A.}~\bibnamefont {Raggio}}, \bibinfo {author} {\bibfnamefont
  {S.}~\bibnamefont {Rinta-Antila}}, \bibinfo {author} {\bibfnamefont
  {J.}~\bibnamefont {Suhonen}}, \bibinfo {author} {\bibfnamefont
  {V.}~\bibnamefont {Virtanen}}, \bibinfo {author} {\bibfnamefont {A.~P.}\
  \bibnamefont {Weaver}}, \bibinfo {author} {\bibfnamefont {A.}~\bibnamefont
  {Zadvornaya}}, \ and\ \bibinfo {author} {\bibfnamefont {A.}~\bibnamefont
  {Jokinen}},\ }\href {\doibase 10.1103/PhysRevC.103.065502} {\bibfield
  {journal} {\bibinfo  {journal} {Phys. Rev. C}\ }\textbf {\bibinfo {volume}
  {103}},\ \bibinfo {pages} {065502} (\bibinfo {year}
  {2021}{\natexlab{a}})}\BibitemShut {NoStop}%
\bibitem [{\citenamefont {Ge}\ \emph {et~al.}(2021{\natexlab{b}})\citenamefont
  {Ge}, \citenamefont {Eronen}, \citenamefont {Tyrin}, \citenamefont {Kotila},
  \citenamefont {Kostensalo}, \citenamefont {Nesterenko}, \citenamefont
  {Beliuskina}, \citenamefont {de~Groote}, \citenamefont {de~Roubin},
  \citenamefont {Geldhof}, \citenamefont {Gins}, \citenamefont {Hukkanen},
  \citenamefont {Jokinen}, \citenamefont {Kankainen}, \citenamefont
  {Koszor\'us}, \citenamefont {Krivoruchenko}, \citenamefont {Kujanp\"a\"a},
  \citenamefont {Moore}, \citenamefont {Raggio}, \citenamefont {Rinta-Antila},
  \citenamefont {Suhonen}, \citenamefont {Virtanen}, \citenamefont {Weaver},\
  and\ \citenamefont {Zadvornaya}}]{Ge2021_159Dy}%
  \BibitemOpen
  \bibfield  {author} {\bibinfo {author} {\bibfnamefont {Z.}~\bibnamefont
  {Ge}}, \bibinfo {author} {\bibfnamefont {T.}~\bibnamefont {Eronen}}, \bibinfo
  {author} {\bibfnamefont {K.~S.}\ \bibnamefont {Tyrin}}, \bibinfo {author}
  {\bibfnamefont {J.}~\bibnamefont {Kotila}}, \bibinfo {author} {\bibfnamefont
  {J.}~\bibnamefont {Kostensalo}}, \bibinfo {author} {\bibfnamefont {D.~A.}\
  \bibnamefont {Nesterenko}}, \bibinfo {author} {\bibfnamefont
  {O.}~\bibnamefont {Beliuskina}}, \bibinfo {author} {\bibfnamefont
  {R.}~\bibnamefont {de~Groote}}, \bibinfo {author} {\bibfnamefont
  {A.}~\bibnamefont {de~Roubin}}, \bibinfo {author} {\bibfnamefont
  {S.}~\bibnamefont {Geldhof}}, \bibinfo {author} {\bibfnamefont
  {W.}~\bibnamefont {Gins}}, \bibinfo {author} {\bibfnamefont {M.}~\bibnamefont
  {Hukkanen}}, \bibinfo {author} {\bibfnamefont {A.}~\bibnamefont {Jokinen}},
  \bibinfo {author} {\bibfnamefont {A.}~\bibnamefont {Kankainen}}, \bibinfo
  {author} {\bibfnamefont {A.}~\bibnamefont {Koszor\'us}}, \bibinfo {author}
  {\bibfnamefont {M.~I.}\ \bibnamefont {Krivoruchenko}}, \bibinfo {author}
  {\bibfnamefont {S.}~\bibnamefont {Kujanp\"a\"a}}, \bibinfo {author}
  {\bibfnamefont {I.~D.}\ \bibnamefont {Moore}}, \bibinfo {author}
  {\bibfnamefont {A.}~\bibnamefont {Raggio}}, \bibinfo {author} {\bibfnamefont
  {S.}~\bibnamefont {Rinta-Antila}}, \bibinfo {author} {\bibfnamefont
  {J.}~\bibnamefont {Suhonen}}, \bibinfo {author} {\bibfnamefont
  {V.}~\bibnamefont {Virtanen}}, \bibinfo {author} {\bibfnamefont {A.~P.}\
  \bibnamefont {Weaver}}, \ and\ \bibinfo {author} {\bibfnamefont
  {A.}~\bibnamefont {Zadvornaya}},\ }\href {\doibase
  10.1103/PhysRevLett.127.272301} {\bibfield  {journal} {\bibinfo  {journal}
  {Phys. Rev. Lett.}\ }\textbf {\bibinfo {volume} {127}},\ \bibinfo {pages}
  {272301} (\bibinfo {year} {2021}{\natexlab{b}})}\BibitemShut {NoStop}%
\bibitem [{\citenamefont {Kankainen}\ \emph {et~al.}(2020)\citenamefont
  {Kankainen}, \citenamefont {Eronen}, \citenamefont {Nesterenko},
  \citenamefont {de~Roubin},\ and\ \citenamefont {Vil\'{e}n}}]{Kan2020}%
  \BibitemOpen
  \bibfield  {author} {\bibinfo {author} {\bibfnamefont {A.}~\bibnamefont
  {Kankainen}}, \bibinfo {author} {\bibfnamefont {T.}~\bibnamefont {Eronen}},
  \bibinfo {author} {\bibfnamefont {D.}~\bibnamefont {Nesterenko}}, \bibinfo
  {author} {\bibfnamefont {A.}~\bibnamefont {de~Roubin}}, \ and\ \bibinfo
  {author} {\bibfnamefont {M.}~\bibnamefont {Vil\'{e}n}},\ }\href {\doibase
  10.1007/s10751-020-01711-5} {\bibfield  {journal} {\bibinfo  {journal}
  {Hyperfine Interactions}\ }\textbf {\bibinfo {volume} {241}},\ \bibinfo
  {pages} {43} (\bibinfo {year} {2020})}\BibitemShut {NoStop}%
\bibitem [{\citenamefont {Savard}\ \emph {et~al.}(2001)\citenamefont {Savard},
  \citenamefont {Barber}, \citenamefont {Boudreau}, \citenamefont {Buchinger},
  \citenamefont {Caggiano}, \citenamefont {Clark}, \citenamefont {Crawford},
  \citenamefont {Fukutani}, \citenamefont {Gulick}, \citenamefont {Hardy} \emph
  {et~al.}}]{savard2001}%
  \BibitemOpen
  \bibfield  {author} {\bibinfo {author} {\bibfnamefont {G.}~\bibnamefont
  {Savard}}, \bibinfo {author} {\bibfnamefont {R.}~\bibnamefont {Barber}},
  \bibinfo {author} {\bibfnamefont {C.}~\bibnamefont {Boudreau}}, \bibinfo
  {author} {\bibfnamefont {F.}~\bibnamefont {Buchinger}}, \bibinfo {author}
  {\bibfnamefont {J.}~\bibnamefont {Caggiano}}, \bibinfo {author}
  {\bibfnamefont {J.}~\bibnamefont {Clark}}, \bibinfo {author} {\bibfnamefont
  {J.}~\bibnamefont {Crawford}}, \bibinfo {author} {\bibfnamefont
  {H.}~\bibnamefont {Fukutani}}, \bibinfo {author} {\bibfnamefont
  {S.}~\bibnamefont {Gulick}}, \bibinfo {author} {\bibfnamefont
  {J.}~\bibnamefont {Hardy}},  \emph {et~al.},\ }\href@noop {} {\bibfield
  {journal} {\bibinfo  {journal} {Hyperfine Interactions}\ }\textbf {\bibinfo
  {volume} {132}},\ \bibinfo {pages} {223} (\bibinfo {year}
  {2001})}\BibitemShut {NoStop}%
\bibitem [{\citenamefont {Orford}\ \emph {et~al.}(2020)\citenamefont {Orford},
  \citenamefont {Clark}, \citenamefont {Savard}, \citenamefont {Aprahamian},
  \citenamefont {Buchinger}, \citenamefont {Burkey}, \citenamefont {Gorelov},
  \citenamefont {Klimes}, \citenamefont {Morgan}, \citenamefont {Nystrom} \emph
  {et~al.}}]{orford2020}%
  \BibitemOpen
  \bibfield  {author} {\bibinfo {author} {\bibfnamefont {R.}~\bibnamefont
  {Orford}}, \bibinfo {author} {\bibfnamefont {J.}~\bibnamefont {Clark}},
  \bibinfo {author} {\bibfnamefont {G.}~\bibnamefont {Savard}}, \bibinfo
  {author} {\bibfnamefont {A.}~\bibnamefont {Aprahamian}}, \bibinfo {author}
  {\bibfnamefont {F.}~\bibnamefont {Buchinger}}, \bibinfo {author}
  {\bibfnamefont {M.}~\bibnamefont {Burkey}}, \bibinfo {author} {\bibfnamefont
  {D.}~\bibnamefont {Gorelov}}, \bibinfo {author} {\bibfnamefont
  {J.}~\bibnamefont {Klimes}}, \bibinfo {author} {\bibfnamefont
  {G.}~\bibnamefont {Morgan}}, \bibinfo {author} {\bibfnamefont
  {A.}~\bibnamefont {Nystrom}},  \emph {et~al.},\ }\href@noop {} {\bibfield
  {journal} {\bibinfo  {journal} {Nuclear Instruments and Methods in Physics
  Research Section B: Beam Interactions with Materials and Atoms}\ }\textbf
  {\bibinfo {volume} {463}},\ \bibinfo {pages} {491} (\bibinfo {year}
  {2020})}\BibitemShut {NoStop}%
\bibitem [{\citenamefont {Savard}\ \emph {et~al.}(2008)\citenamefont {Savard},
  \citenamefont {Baker}, \citenamefont {Davids}, \citenamefont {Levand},
  \citenamefont {Moore}, \citenamefont {Pardo}, \citenamefont {Vondrasek},
  \citenamefont {Zabransky},\ and\ \citenamefont {Zinkann}}]{savard2008}%
  \BibitemOpen
  \bibfield  {author} {\bibinfo {author} {\bibfnamefont {G.}~\bibnamefont
  {Savard}}, \bibinfo {author} {\bibfnamefont {S.}~\bibnamefont {Baker}},
  \bibinfo {author} {\bibfnamefont {C.}~\bibnamefont {Davids}}, \bibinfo
  {author} {\bibfnamefont {A.}~\bibnamefont {Levand}}, \bibinfo {author}
  {\bibfnamefont {E.}~\bibnamefont {Moore}}, \bibinfo {author} {\bibfnamefont
  {R.}~\bibnamefont {Pardo}}, \bibinfo {author} {\bibfnamefont
  {R.}~\bibnamefont {Vondrasek}}, \bibinfo {author} {\bibfnamefont
  {B.}~\bibnamefont {Zabransky}}, \ and\ \bibinfo {author} {\bibfnamefont
  {G.}~\bibnamefont {Zinkann}},\ }\href@noop {} {\bibfield  {journal} {\bibinfo
   {journal} {Nuclear Instruments and Methods in Physics Research Section B:
  Beam Interactions with Materials and Atoms}\ }\textbf {\bibinfo {volume}
  {266}},\ \bibinfo {pages} {4086} (\bibinfo {year} {2008})}\BibitemShut
  {NoStop}%
\bibitem [{\citenamefont {Savard}\ \emph {et~al.}(2003)\citenamefont {Savard},
  \citenamefont {Clark}, \citenamefont {Boudreau}, \citenamefont {Buchinger},
  \citenamefont {Crawford}, \citenamefont {Geissel}, \citenamefont {Greene},
  \citenamefont {Gulick}, \citenamefont {Heinz}, \citenamefont {Lee},
  \citenamefont {Levand}, \citenamefont {Maier}, \citenamefont
  {M{\"u}nzenberg}, \citenamefont {Scheidenberger}, \citenamefont {Seweryniak},
  \citenamefont {Sharma}, \citenamefont {Sprouse}, \citenamefont {Vaz},
  \citenamefont {Wang}, \citenamefont {Zabransky},\ and\ \citenamefont
  {Zhou}}]{SAVARD2003}%
  \BibitemOpen
  \bibfield  {author} {\bibinfo {author} {\bibfnamefont {G.}~\bibnamefont
  {Savard}}, \bibinfo {author} {\bibfnamefont {J.}~\bibnamefont {Clark}},
  \bibinfo {author} {\bibfnamefont {C.}~\bibnamefont {Boudreau}}, \bibinfo
  {author} {\bibfnamefont {F.}~\bibnamefont {Buchinger}}, \bibinfo {author}
  {\bibfnamefont {J.}~\bibnamefont {Crawford}}, \bibinfo {author}
  {\bibfnamefont {H.}~\bibnamefont {Geissel}}, \bibinfo {author} {\bibfnamefont
  {J.}~\bibnamefont {Greene}}, \bibinfo {author} {\bibfnamefont
  {S.}~\bibnamefont {Gulick}}, \bibinfo {author} {\bibfnamefont
  {A.}~\bibnamefont {Heinz}}, \bibinfo {author} {\bibfnamefont
  {J.}~\bibnamefont {Lee}}, \bibinfo {author} {\bibfnamefont {A.}~\bibnamefont
  {Levand}}, \bibinfo {author} {\bibfnamefont {M.}~\bibnamefont {Maier}},
  \bibinfo {author} {\bibfnamefont {G.}~\bibnamefont {M{\"u}nzenberg}},
  \bibinfo {author} {\bibfnamefont {C.}~\bibnamefont {Scheidenberger}},
  \bibinfo {author} {\bibfnamefont {D.}~\bibnamefont {Seweryniak}}, \bibinfo
  {author} {\bibfnamefont {K.}~\bibnamefont {Sharma}}, \bibinfo {author}
  {\bibfnamefont {G.}~\bibnamefont {Sprouse}}, \bibinfo {author} {\bibfnamefont
  {J.}~\bibnamefont {Vaz}}, \bibinfo {author} {\bibfnamefont {J.}~\bibnamefont
  {Wang}}, \bibinfo {author} {\bibfnamefont {B.}~\bibnamefont {Zabransky}}, \
  and\ \bibinfo {author} {\bibfnamefont {Z.}~\bibnamefont {Zhou}},\ }\href@noop
  {} {\bibfield  {journal} {\bibinfo  {journal} {Nuclear Instruments and
  Methods in Physics Research Section B: Beam Interactions with Materials and
  Atoms}\ }\textbf {\bibinfo {volume} {204}},\ \bibinfo {pages} {582} (\bibinfo
  {year} {2003})}\BibitemShut {NoStop}%
\bibitem [{\citenamefont {Hirsh}\ \emph {et~al.}(2016)\citenamefont {Hirsh},
  \citenamefont {Paul}, \citenamefont {Burkey}, \citenamefont {Aprahamian},
  \citenamefont {Buchinger}, \citenamefont {Caldwell}, \citenamefont {Clark},
  \citenamefont {Levand}, \citenamefont {Ying}, \citenamefont {Marley},
  \citenamefont {Morgan}, \citenamefont {Nystrom}, \citenamefont {Orford},
  \citenamefont {Galván}, \citenamefont {Rohrer}, \citenamefont {Savard},
  \citenamefont {Sharma},\ and\ \citenamefont {Siegl}}]{HIRSH2016}%
  \BibitemOpen
  \bibfield  {author} {\bibinfo {author} {\bibfnamefont {T.~Y.}\ \bibnamefont
  {Hirsh}}, \bibinfo {author} {\bibfnamefont {N.}~\bibnamefont {Paul}},
  \bibinfo {author} {\bibfnamefont {M.}~\bibnamefont {Burkey}}, \bibinfo
  {author} {\bibfnamefont {A.}~\bibnamefont {Aprahamian}}, \bibinfo {author}
  {\bibfnamefont {F.}~\bibnamefont {Buchinger}}, \bibinfo {author}
  {\bibfnamefont {S.}~\bibnamefont {Caldwell}}, \bibinfo {author}
  {\bibfnamefont {J.~A.}\ \bibnamefont {Clark}}, \bibinfo {author}
  {\bibfnamefont {A.~F.}\ \bibnamefont {Levand}}, \bibinfo {author}
  {\bibfnamefont {L.~L.}\ \bibnamefont {Ying}}, \bibinfo {author}
  {\bibfnamefont {S.~T.}\ \bibnamefont {Marley}}, \bibinfo {author}
  {\bibfnamefont {G.~E.}\ \bibnamefont {Morgan}}, \bibinfo {author}
  {\bibfnamefont {A.}~\bibnamefont {Nystrom}}, \bibinfo {author} {\bibfnamefont
  {R.}~\bibnamefont {Orford}}, \bibinfo {author} {\bibfnamefont {A.~P.}\
  \bibnamefont {Galván}}, \bibinfo {author} {\bibfnamefont {J.}~\bibnamefont
  {Rohrer}}, \bibinfo {author} {\bibfnamefont {G.}~\bibnamefont {Savard}},
  \bibinfo {author} {\bibfnamefont {K.~S.}\ \bibnamefont {Sharma}}, \ and\
  \bibinfo {author} {\bibfnamefont {K.}~\bibnamefont {Siegl}},\ }\href
  {\doibase https://doi.org/10.1016/j.nimb.2015.12.037} {\bibfield  {journal}
  {\bibinfo  {journal} {Nuclear Instruments and Methods in Physics Research
  Section B: Beam Interactions with Materials and Atoms}\ }\textbf {\bibinfo
  {volume} {376}},\ \bibinfo {pages} {229} (\bibinfo {year} {2016})},\ \bibinfo
  {note} {proceedings of the XVIIth International Conference on Electromagnetic
  Isotope Separators and Related Topics (EMIS2015), Grand Rapids, MI, U.S.A.,
  11-15 May 2015}\BibitemShut {NoStop}%
\bibitem [{\citenamefont {Bradbury}\ and\ \citenamefont
  {Nielsen}(1936)}]{BNG1936}%
  \BibitemOpen
  \bibfield  {author} {\bibinfo {author} {\bibfnamefont {N.~E.}\ \bibnamefont
  {Bradbury}}\ and\ \bibinfo {author} {\bibfnamefont {R.~A.}\ \bibnamefont
  {Nielsen}},\ }\href {\doibase 10.1103/PhysRev.49.388} {\bibfield  {journal}
  {\bibinfo  {journal} {Phys. Rev.}\ }\textbf {\bibinfo {volume} {49}},\
  \bibinfo {pages} {388} (\bibinfo {year} {1936})}\BibitemShut {NoStop}%
\bibitem [{\citenamefont {Brown}\ and\ \citenamefont
  {Gabrielse}(1986)}]{Brown1986}%
  \BibitemOpen
  \bibfield  {author} {\bibinfo {author} {\bibfnamefont {L.~S.}\ \bibnamefont
  {Brown}}\ and\ \bibinfo {author} {\bibfnamefont {G.}~\bibnamefont
  {Gabrielse}},\ }\href {\doibase 10.1103/RevModPhys.58.233} {\bibfield
  {journal} {\bibinfo  {journal} {Rev. Mod. Phys.}\ }\textbf {\bibinfo {volume}
  {58}},\ \bibinfo {pages} {233} (\bibinfo {year} {1986})}\BibitemShut
  {NoStop}%
\bibitem [{\citenamefont {Eliseev}\ \emph {et~al.}(2013)\citenamefont
  {Eliseev}, \citenamefont {Blaum}, \citenamefont {Block}, \citenamefont
  {Droese}, \citenamefont {Goncharov}, \citenamefont {Minaya~Ramirez},
  \citenamefont {Nesterenko}, \citenamefont {Novikov},\ and\ \citenamefont
  {Schweikhard}}]{eliseev2013}%
  \BibitemOpen
  \bibfield  {author} {\bibinfo {author} {\bibfnamefont {S.}~\bibnamefont
  {Eliseev}}, \bibinfo {author} {\bibfnamefont {K.}~\bibnamefont {Blaum}},
  \bibinfo {author} {\bibfnamefont {M.}~\bibnamefont {Block}}, \bibinfo
  {author} {\bibfnamefont {C.}~\bibnamefont {Droese}}, \bibinfo {author}
  {\bibfnamefont {M.}~\bibnamefont {Goncharov}}, \bibinfo {author}
  {\bibfnamefont {E.}~\bibnamefont {Minaya~Ramirez}}, \bibinfo {author}
  {\bibfnamefont {D.}~\bibnamefont {Nesterenko}}, \bibinfo {author}
  {\bibfnamefont {Y.~N.}\ \bibnamefont {Novikov}}, \ and\ \bibinfo {author}
  {\bibfnamefont {L.}~\bibnamefont {Schweikhard}},\ }\href@noop {} {\bibfield
  {journal} {\bibinfo  {journal} {Physical Review Letters}\ }\textbf {\bibinfo
  {volume} {110}},\ \bibinfo {pages} {082501} (\bibinfo {year}
  {2013})}\BibitemShut {NoStop}%
\bibitem [{\citenamefont {Eliseev}\ \emph {et~al.}(2014)\citenamefont
  {Eliseev}, \citenamefont {Blaum}, \citenamefont {Block}, \citenamefont
  {D{\"o}rr}, \citenamefont {Droese}, \citenamefont {Eronen}, \citenamefont
  {Goncharov}, \citenamefont {H{\"o}cker}, \citenamefont {Ketter},
  \citenamefont {Ramirez} \emph {et~al.}}]{eliseev2014}%
  \BibitemOpen
  \bibfield  {author} {\bibinfo {author} {\bibfnamefont {S.}~\bibnamefont
  {Eliseev}}, \bibinfo {author} {\bibfnamefont {K.}~\bibnamefont {Blaum}},
  \bibinfo {author} {\bibfnamefont {M.}~\bibnamefont {Block}}, \bibinfo
  {author} {\bibfnamefont {A.}~\bibnamefont {D{\"o}rr}}, \bibinfo {author}
  {\bibfnamefont {C.}~\bibnamefont {Droese}}, \bibinfo {author} {\bibfnamefont
  {T.}~\bibnamefont {Eronen}}, \bibinfo {author} {\bibfnamefont
  {M.}~\bibnamefont {Goncharov}}, \bibinfo {author} {\bibfnamefont
  {M.}~\bibnamefont {H{\"o}cker}}, \bibinfo {author} {\bibfnamefont
  {J.}~\bibnamefont {Ketter}}, \bibinfo {author} {\bibfnamefont {E.~M.}\
  \bibnamefont {Ramirez}},  \emph {et~al.},\ }\href@noop {} {\bibfield
  {journal} {\bibinfo  {journal} {Applied Physics B}\ }\textbf {\bibinfo
  {volume} {114}},\ \bibinfo {pages} {107} (\bibinfo {year}
  {2014})}\BibitemShut {NoStop}%
\bibitem [{\citenamefont {Weber}\ \emph {et~al.}(2022)\citenamefont {Weber},
  \citenamefont {Ray}, \citenamefont {Valverde}, \citenamefont {Clark},\ and\
  \citenamefont {Sharma}}]{Weber2022}%
  \BibitemOpen
  \bibfield  {author} {\bibinfo {author} {\bibfnamefont {C.}~\bibnamefont
  {Weber}}, \bibinfo {author} {\bibfnamefont {D.}~\bibnamefont {Ray}}, \bibinfo
  {author} {\bibfnamefont {A.}~\bibnamefont {Valverde}}, \bibinfo {author}
  {\bibfnamefont {J.}~\bibnamefont {Clark}}, \ and\ \bibinfo {author}
  {\bibfnamefont {K.}~\bibnamefont {Sharma}},\ }\href@noop {} {\bibfield
  {journal} {\bibinfo  {journal} {Nuclear Instruments and Methods in Physics
  Research Section A: Accelerators, Spectrometers, Detectors and Associated
  Equipment}\ }\textbf {\bibinfo {volume} {1027}},\ \bibinfo {pages} {166299}
  (\bibinfo {year} {2022})}\BibitemShut {NoStop}%
\bibitem [{\citenamefont {Rahaman}\ \emph {et~al.}(2009)\citenamefont
  {Rahaman}, \citenamefont {Elomaa}, \citenamefont {Eronen}, \citenamefont
  {Hakala}, \citenamefont {Jokinen}, \citenamefont {Kankainen}, \citenamefont
  {Rissanen}, \citenamefont {Suhonen}, \citenamefont {Weber},\ and\
  \citenamefont {\"Ayst\"o}}]{Rahaman2009}%
  \BibitemOpen
  \bibfield  {author} {\bibinfo {author} {\bibfnamefont {S.}~\bibnamefont
  {Rahaman}}, \bibinfo {author} {\bibfnamefont {V.-V.}\ \bibnamefont {Elomaa}},
  \bibinfo {author} {\bibfnamefont {T.}~\bibnamefont {Eronen}}, \bibinfo
  {author} {\bibfnamefont {J.}~\bibnamefont {Hakala}}, \bibinfo {author}
  {\bibfnamefont {A.}~\bibnamefont {Jokinen}}, \bibinfo {author} {\bibfnamefont
  {A.}~\bibnamefont {Kankainen}}, \bibinfo {author} {\bibfnamefont
  {J.}~\bibnamefont {Rissanen}}, \bibinfo {author} {\bibfnamefont
  {J.}~\bibnamefont {Suhonen}}, \bibinfo {author} {\bibfnamefont
  {C.}~\bibnamefont {Weber}}, \ and\ \bibinfo {author} {\bibfnamefont
  {J.}~\bibnamefont {\"Ayst\"o}},\ }\href {\doibase
  10.1103/PhysRevLett.103.042501} {\bibfield  {journal} {\bibinfo  {journal}
  {Phys. Rev. Lett.}\ }\textbf {\bibinfo {volume} {103}},\ \bibinfo {pages}
  {042501} (\bibinfo {year} {2009})}\BibitemShut {NoStop}%
\bibitem [{\citenamefont {Kondev}\ \emph {et~al.}(2021)\citenamefont {Kondev},
  \citenamefont {Wang}, \citenamefont {Huang}, \citenamefont {Naimi},\ and\
  \citenamefont {Audi}}]{Kondev2021}%
  \BibitemOpen
  \bibfield  {author} {\bibinfo {author} {\bibfnamefont {F.}~\bibnamefont
  {Kondev}}, \bibinfo {author} {\bibfnamefont {M.}~\bibnamefont {Wang}},
  \bibinfo {author} {\bibfnamefont {W.}~\bibnamefont {Huang}}, \bibinfo
  {author} {\bibfnamefont {S.}~\bibnamefont {Naimi}}, \ and\ \bibinfo {author}
  {\bibfnamefont {G.}~\bibnamefont {Audi}},\ }\href {\doibase
  10.1088/1674-1137/abddae} {\bibfield  {journal} {\bibinfo  {journal} {Chinese
  Physics C}\ }\textbf {\bibinfo {volume} {45}},\ \bibinfo {pages} {030001}
  (\bibinfo {year} {2021})}\BibitemShut {NoStop}%
\bibitem [{\citenamefont {Tiesinga}\ \emph {et~al.}(2021)\citenamefont
  {Tiesinga}, \citenamefont {Mohr}, \citenamefont {Newell},\ and\ \citenamefont
  {Taylor}}]{CODATA2018}%
  \BibitemOpen
  \bibfield  {author} {\bibinfo {author} {\bibfnamefont {E.}~\bibnamefont
  {Tiesinga}}, \bibinfo {author} {\bibfnamefont {P.~J.}\ \bibnamefont {Mohr}},
  \bibinfo {author} {\bibfnamefont {D.~B.}\ \bibnamefont {Newell}}, \ and\
  \bibinfo {author} {\bibfnamefont {B.~N.}\ \bibnamefont {Taylor}},\ }\href
  {\doibase 10.1103/RevModPhys.93.025010} {\bibfield  {journal} {\bibinfo
  {journal} {Rev. Mod. Phys.}\ }\textbf {\bibinfo {volume} {93}},\ \bibinfo
  {pages} {025010} (\bibinfo {year} {2021})}\BibitemShut {NoStop}%
\bibitem [{nnd()}]{nndc}%
  \BibitemOpen
  \href@noop {} {\enquote {\bibinfo {title} {National nuclear data center},}\
  }\bibinfo {howpublished} {\url{https://www.nndc.bnl.gov/}}\BibitemShut
  {NoStop}%
\bibitem [{\citenamefont {Matumoto}\ and\ \citenamefont
  {Tamura}(1970)}]{Matumoto1970}%
  \BibitemOpen
  \bibfield  {author} {\bibinfo {author} {\bibfnamefont {Z.-I.}\ \bibnamefont
  {Matumoto}}\ and\ \bibinfo {author} {\bibfnamefont {T.}~\bibnamefont
  {Tamura}},\ }\href {\doibase 10.1143/JPSJ.29.1116} {\bibfield  {journal}
  {\bibinfo  {journal} {Journal of the Physical Society of Japan}\ }\textbf
  {\bibinfo {volume} {29}},\ \bibinfo {pages} {1116} (\bibinfo {year}
  {1970})}\BibitemShut {NoStop}%
\bibitem [{\citenamefont {Fogelberg}\ \emph {et~al.}(1990)\citenamefont
  {Fogelberg}, \citenamefont {Zongyuan}, \citenamefont {Ekström},
  \citenamefont {Lund}, \citenamefont {Aleklett},\ and\ \citenamefont
  {Sihver}}]{Fogelberg1990}%
  \BibitemOpen
  \bibfield  {author} {\bibinfo {author} {\bibfnamefont {B.}~\bibnamefont
  {Fogelberg}}, \bibinfo {author} {\bibfnamefont {Y.}~\bibnamefont {Zongyuan}},
  \bibinfo {author} {\bibfnamefont {B.}~\bibnamefont {Ekström}}, \bibinfo
  {author} {\bibfnamefont {E.}~\bibnamefont {Lund}}, \bibinfo {author}
  {\bibfnamefont {K.}~\bibnamefont {Aleklett}}, \ and\ \bibinfo {author}
  {\bibfnamefont {L.}~\bibnamefont {Sihver}},\ }\href {\doibase
  10.1007/BF01289690} {\bibfield  {journal} {\bibinfo  {journal} {Zeitschrift
  für Physik A Atomic Nuclei}\ }\textbf {\bibinfo {volume} {337}},\ \bibinfo
  {pages} {251} (\bibinfo {year} {1990})}\BibitemShut {NoStop}%
\bibitem [{\citenamefont {Breitenfeldt}\ \emph {et~al.}(2010)\citenamefont
  {Breitenfeldt}, \citenamefont {Borgmann}, \citenamefont {Audi}, \citenamefont
  {Baruah}, \citenamefont {Beck}, \citenamefont {Blaum}, \citenamefont
  {B\"ohm}, \citenamefont {Cakirli}, \citenamefont {Casten}, \citenamefont
  {Delahaye}, \citenamefont {Dworschak}, \citenamefont {George}, \citenamefont
  {Herfurth}, \citenamefont {Herlert}, \citenamefont {Kellerbauer},
  \citenamefont {Kowalska}, \citenamefont {Lunney}, \citenamefont
  {Minaya-Ramirez}, \citenamefont {Naimi}, \citenamefont {Neidherr},
  \citenamefont {Rosenbusch}, \citenamefont {Savreux}, \citenamefont {Schwarz},
  \citenamefont {Schweikhard},\ and\ \citenamefont
  {Yazidjian}}]{Breitenfeldt2010}%
  \BibitemOpen
  \bibfield  {author} {\bibinfo {author} {\bibfnamefont {M.}~\bibnamefont
  {Breitenfeldt}}, \bibinfo {author} {\bibfnamefont {C.}~\bibnamefont
  {Borgmann}}, \bibinfo {author} {\bibfnamefont {G.}~\bibnamefont {Audi}},
  \bibinfo {author} {\bibfnamefont {S.}~\bibnamefont {Baruah}}, \bibinfo
  {author} {\bibfnamefont {D.}~\bibnamefont {Beck}}, \bibinfo {author}
  {\bibfnamefont {K.}~\bibnamefont {Blaum}}, \bibinfo {author} {\bibfnamefont
  {C.}~\bibnamefont {B\"ohm}}, \bibinfo {author} {\bibfnamefont {R.~B.}\
  \bibnamefont {Cakirli}}, \bibinfo {author} {\bibfnamefont {R.~F.}\
  \bibnamefont {Casten}}, \bibinfo {author} {\bibfnamefont {P.}~\bibnamefont
  {Delahaye}}, \bibinfo {author} {\bibfnamefont {M.}~\bibnamefont {Dworschak}},
  \bibinfo {author} {\bibfnamefont {S.}~\bibnamefont {George}}, \bibinfo
  {author} {\bibfnamefont {F.}~\bibnamefont {Herfurth}}, \bibinfo {author}
  {\bibfnamefont {A.}~\bibnamefont {Herlert}}, \bibinfo {author} {\bibfnamefont
  {A.}~\bibnamefont {Kellerbauer}}, \bibinfo {author} {\bibfnamefont
  {M.}~\bibnamefont {Kowalska}}, \bibinfo {author} {\bibfnamefont
  {D.}~\bibnamefont {Lunney}}, \bibinfo {author} {\bibfnamefont
  {E.}~\bibnamefont {Minaya-Ramirez}}, \bibinfo {author} {\bibfnamefont
  {S.}~\bibnamefont {Naimi}}, \bibinfo {author} {\bibfnamefont
  {D.}~\bibnamefont {Neidherr}}, \bibinfo {author} {\bibfnamefont
  {M.}~\bibnamefont {Rosenbusch}}, \bibinfo {author} {\bibfnamefont
  {R.}~\bibnamefont {Savreux}}, \bibinfo {author} {\bibfnamefont
  {S.}~\bibnamefont {Schwarz}}, \bibinfo {author} {\bibfnamefont
  {L.}~\bibnamefont {Schweikhard}}, \ and\ \bibinfo {author} {\bibfnamefont
  {C.}~\bibnamefont {Yazidjian}},\ }\href {\doibase 10.1103/PhysRevC.81.034313}
  {\bibfield  {journal} {\bibinfo  {journal} {Phys. Rev. C}\ }\textbf {\bibinfo
  {volume} {81}},\ \bibinfo {pages} {034313} (\bibinfo {year}
  {2010})}\BibitemShut {NoStop}%
\end{thebibliography}%

\end{document}